\documentclass[%
 aip,
 amsmath,amssymb,
 reprint,%
]{revtex4-2}

\usepackage{graphicx}
\usepackage{dcolumn}
\usepackage{bm}

\usepackage[utf8]{inputenc}
\usepackage[T1]{fontenc}
\usepackage{mathptmx}

\usepackage{braket}
\usepackage{amsmath,amssymb}
\usepackage{physics}
\usepackage{amsmath}
\usepackage{bm}
\usepackage{bbold}
\usepackage{braket}
\usepackage{xcolor}
\usepackage{placeins}
\usepackage{ulem}
\usepackage{natbib}

\begin{document}


\title[]{Semiclassical Dynamics in Wigner Phase Space II : nonadiabatic Hybrid Wigner Dynamics}

\author{Shreyas Malpathak}
\altaffiliation[Current affiliation: ]{Department of Physical and Environmental Sciences, University of Toronto Scarborough, Toronto, Ontario M1C 1A4, Canada}
\author{Nandini Ananth}%
 \email{ananth@cornell.edu}
\affiliation{Department of Chemistry and Chemical Biology, Baker Laboratory, Cornell University
Ithaca,14853 NY, USA}%


\date{\today}

\begin{abstract}
 We present an approximate semiclassical (SC) framework for mixed quantized dynamics in Wigner phase space in a two-part series. In the first article, we introduced the Adiabatic Hybrid Wigner Dynamics (AHWD) method that allows for a few important `system' degrees of freedom to be quantized using high-level Double Herman-Kluk SC theory while describing the rest (the `bath') using classical-limit Linearized SC theory. In this second article, we extend our hybrid Wigner dynamics to nonadiabatic processes. The resulting Nonadiabatic Hybrid Wigner Dynamics (NHWD) has two variants that differ in the choice of degrees of freedom to be quantized. Specifically, we introduce NHWD(E) where only the electronic state variables are quantized and the NHWD(V) where both electronic state variables and a handful of strongly coupled nuclear modes are quantized. We show that while NHWD(E) proves accurate for a wide range of scattering models and spin-boson models, systems where a few nuclear modes are strongly coupled to electronic states require NHWD(V) to accurately capture the long-time dynamics. Taken together, we show that AHWD and NHWD represent a new framework for SC simulations of high-dimensional systems with significant quantum effects.
\end{abstract}

\maketitle

\section{Introduction}
Nonadiabatic quantum dynamics have been at the center of many 
studies of chemical phenomena in the last couple of decades, including, but not limited to, exciton energy transfer in photosynthetic protein complexes and molecular aggregates,\cite{Lee2016,Cao2020,Kundu2022a,Kundu2022b,Kundu2022c} the role of electron hole pair excitations in energy relaxation of molecules at metal surfaces,\cite{Golibrzuch2015,Wodtke2016,Rittmeyer2018,Park2019,Dou2020,Zhou2022} electron transfer reactions in biologically important systems,\cite{Blumberger2015} and dynamics through conical intersections.\cite{Domcke2012,Schuurman2018}
Despite tremendous progress in numerically exact quantum dynamic simulation methods like the Quasi-Adiabatic Path Integral (QUAPI),\cite{Makri1992} Hierarchy Equations of Motion (HEOM),\cite{Tanimura1989} Multi-Configuration Time Dependent Hartree (MCTDH),\cite{Meyer1990,Manthe1992,Beck2000} and Tensor-Train Split Operator Fourier Transform (TT-SOFT),\cite{Greene2017} their expensive scaling with respect to the number of degrees of freedom (dofs) limits their use in complex chemical system simulations. 

A wide array of approximate methods with varying accuracy and computational cost have been developed, and a summary can be found in Ref.\citenum{Otero2018}. The \textit{ab initio} Multiple Spawning method (AIMS) propagates nuclear wavepackets represented using a Gaussian basis that is allowed to spawn new basis functions in regions of significant nonadiabatic coupling.\cite{ben-nun2000,Curchod2018} Multi-state imaginary-time path integral methods exploit classical trajectories to capture approximate quantum dynamics in nonadiabatic systems.\cite{liao2002,ananth_2013,richardson2013,chowdhury2017,Ananth2022} Approximate mixed quantum-classical (MQC) methods that treat the nuclear dofs classically include Fewest Switches Surface Hopping (FSSH) approach that allows stochastic 'hopping' between electronic states,\cite{Tully1990,Tully1998} and the Ehrenfest approach, a mean-field method where the force felt by the nuclear dofs is a weighted by the populations of the electronic states.\cite{Tully1998} Nuclear tunneling may be included in the Ehrenfest approach using the army ants tunneling method.\cite{zheng2014}

Semiclassical methods based on the mapping Hamiltonian have emerged as an alternative to mixed quantum-classical methods since they offer a uniform dynamic framework for the treatment of electronic and nuclear dofs.\cite{Stock2005,Miller2009,Liu2021b,Runeson2022} They rely on mapping the discrete electronic dofs to continuous variables, such that exact quantum dynamics of the two coincide. One of the primary features of the mapping approach is its ability to facilitate the extension of a well-established hierarchy of semiclassical dynamic (SC) approximations to nonadiabatic dynamics. 
When used with the Meyer-Miller-Stock-Thoss (MMST) mapping,\cite{Meyer1979,Stock1997} sophisticated SC methods like the Herman-Kluk (HK) propagator\cite{Herman1984} have been  shown to be very accurate for nonadiabatic problems.\cite{Sun1997,Batista1998,Thoss1999,Coronado2000,Guallar2000,Bonella2001a,Bonella2001b,Ananth2007} However, they are prohibitively expensive for even relatively low-dimensional systems, and become 
computationally feasible only when combined with phase-filtering techniques.\cite{Church2018,Malpathak2022} 
Classical-limit SC methods like Linearized SC (LSC)~\cite{Sun1998b} offer a numerically practical alternative, and have found tremendous success in model system studies~\cite{Sun1998b,Wang1999,Rabani1999,Ananth2007,Tao2010,Church2018,Liu2020,Gao2020,Malpathak2024a} and in \textit{ab-initio} simulations.\cite{Tang2019,Hu2021,Weight2021,Talbot2022,Talbot2023,Miyazaki2023} Furthermore, methods based on an alternate mapping schemes,~\cite{Liu2021b,Runeson2019,Runeson2020} have been developed
and show some improvement in accuracy over their MMST counterparts.
\cite{Runeson2019,Runeson2020,Mannouch2020,Mannouch2020b,Runeson2022} 

Despite the success of classical-limit SC methods, avenues to include quantization beyond the classical limit are desired to be able to capture interference effects, and to overcome deficiencies such as the inverted potential problem,\cite{Bonella2001a,Bonella2001b} and the inability to predict wavepacket branching.\cite{Ananth2007,Church2018} The  Partial Linearized Density Matrix (PLDM) method,\cite{Huo2011,Huo2012} and the Forward-Backward Trajectory Solution (FBTS),\cite{Hsieh2012,Hsieh2013} both of which are approximately connected to the Quantum Classical Liouville Equation (QCLE),\cite{Hsieh2012} attempt to improve on classical-limit methods by allowing for a forward-backward trajectory structure for the mapped electronic dofs. However, a theoretical connection between these methods and a full double Herman-Kluk (DHK) approximation for the nonadiabatic correlation function has not been established. Establishing this relationship would not only provide a complete understanding of these approximations but also allow for systematic improvements to incorporate quantized nuclear modes and facilitate investigations into the role of nuclear quantum effect in vibronic systems. 

\begin{figure}
    \centering
    \includegraphics[width=0.48\textwidth]{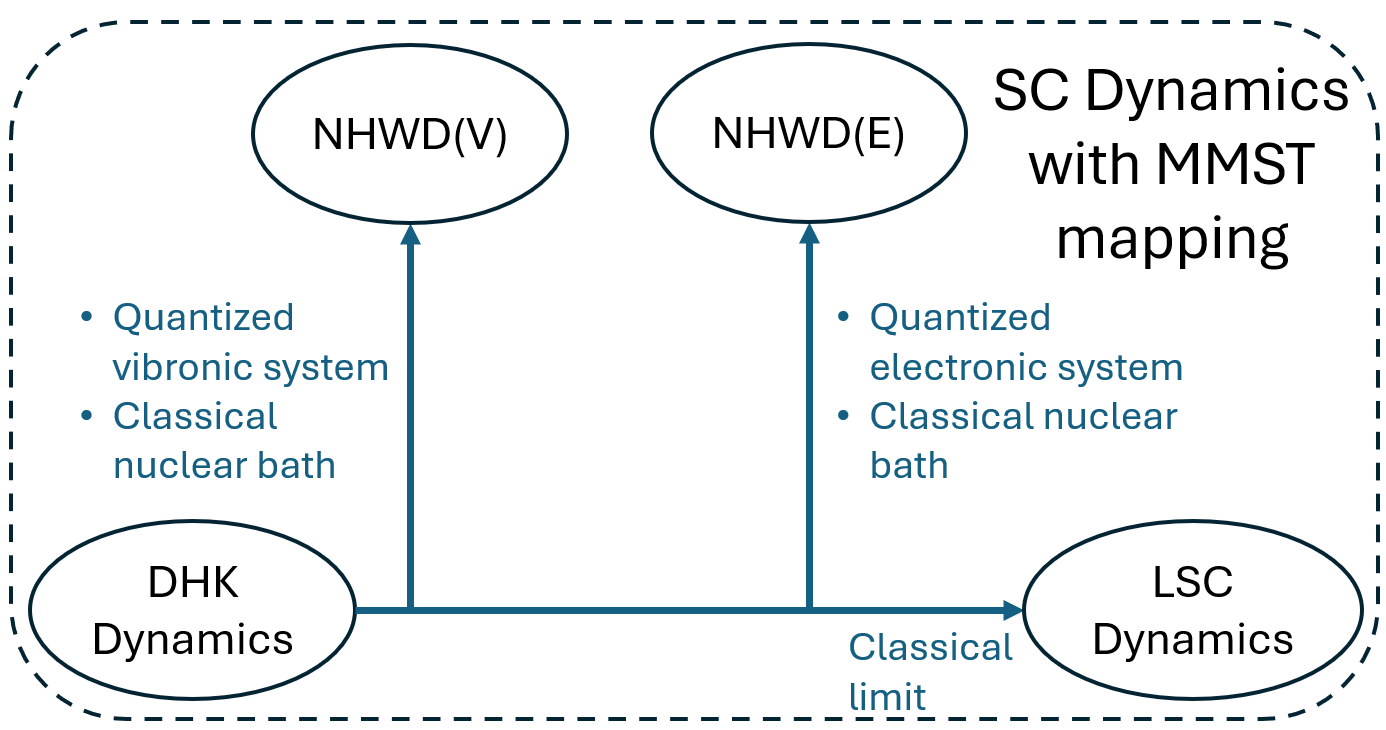}
    \caption{A schematic of the hierarchy of nonadiabatic SC methods with MMST mapping described in this article. Refer to the text for details on the abbreviations used in the schematic.}
    \label{fig:na_schematic}
\end{figure}

In the first article of this series\cite{Malpathak2024b}, hereon referred to as Paper I, we introduced a novel SC method for mixed quantized dynamics in adiabatic systems, the Adiabatic Hybrid Wigner dynamics (AHWD) approach. The AHWD approach allows for a handful of system dofs to be treated at the DHK level of theory that capture almost all quantum effects including coherence effects while the remaining dofs are described using LSC theory, a classical-limit dynamics. In Paper I, we demonstrate that the AHWD offers a path to high accuracy quantum dynamic simulations at a significantly reduced computational cost by mitigating the sign problem common to high-level SC theories like DHK. In this article, we introduce the Nonadiabatic HWD (NHWD) approach that extends the mixed quantized framework to nonadiabatic processes. In choosing the dofs to be quantized, for nonadiabatic system studies, MQC methods typically quantize the electronic state variables while treating the nuclei classically. Here, we start with a similar approximation by partitioning the problem into an electronic `system' to be treated at the DHK level and a nuclear `bath' to be treated at the LSC level.
The resultisg NHWD(E) expression features forward-backward trajectories for the mapped electronic dofs, similar to the PLDM and FBTS. Exploiting the flexibility of the hybrid Wigner dynamics framework, we also introduce NHWD(V) that allows for the quantization of a vibronic `system' consisting of a select few important nuclear dofs and the electronic dofs, while treating the rest of the nuclear dofs in the classical-limit. Both these methods offer a way to systematically improve on classical-limit SC methods, while maintaining a connection with the more sophisticated DHK correlation function. A schematic of the methods discussed in this article is presented in Fig.~\ref{fig:na_schematic}.

The outline of the article is as follows. In Sec.~\ref{sec:theory} we present an alternate avenue to the Double-Herman Kluk (DHK) method in Wigner phase (hereon referred to as Wigner DHK) for nonadiabatic systems described by the MMST Hamiltonian. We then perform a partial stationary phase approximation to derive the mixed quantized NHWD(E) and NHWD(V) expressions. Sec.~\ref{sec:simulation} describes details of the models used to test the newly introduced NHWD methods, along with details of the simulations. In Sec.~\ref{sec:results} the two methods are tested against standard scattering models and spin-boson models parameterized for a variety of regimes. Sec.~\ref{sec:conclusion} concludes.
 
\section{Theory} \label{sec:theory}
\subsection{Wigner DHK with MMST Mapping}

Consider the Hamiltonian,
\begin{align}
    H(\bm{Z}) = \frac{1}{2}\bm{P}^T\cdot\bm{m}^{-1}\cdot\bm{P} + U\left(\bm{X}\right) + \bm{V}\left(\bm{X}\right),
\end{align}
where $\bm{Z}^T\equiv\left(\bm{X},\bm{P}\right)^T$ is the $2D$ dimensional nuclear phase space coordinate, and $\bm{X}$ and $\bm{P}$ are $D$-dimensional vectors of nuclear positions and momenta respectively, $\bm{m}$ is a diagonal matrix of nuclear masses, $U\left(\bm{X}\right)$ is the state-independent potential energy, and $\bm{V}\left(\bm{X}\right)$ is the $F \times F$ dimensional diabatic potential energy matrix with the diabatic potential energy terms on the diagonal and their coupling as the off-diagonal matrix elements. Using the MMST mapping,\cite{Meyer1979,Stock1997} it can be mapped onto the Hamiltonian,
\begin{align}
    H_{MMST}(\bm{\mathsf{Z}}) = \frac{1}{2}\bm{P}^T\cdot\bm{m}^{-1}\cdot\bm{P} + \mathrm{h}_{el}(\bm{z},\bm{X}), \label{eq:h_mmst}
\end{align}
where, $\bm{z^T}\equiv\left(\bm{x},\bm{p}\right)^T$  is the $2F$ dimensional electronic phase space variable, with $\bm{x}$ and $\bm{p}$ being $F$-dimensional vectors of Cartesian position and momenta corresponding to the mapped electronic states. $\bm{\mathsf{X}}^T \equiv \left(\bm{x},\bm{X}\right)^T$ and $\bm{\mathsf{P}}^T \equiv \left(\bm{p},\bm{P}\right)^T$  are the full $N$-dimensional set of positions and momenta, where $N = F + D$, making up the phase space vector $\bm{\mathsf{Z}}^T\equiv\left(\bm{\mathsf{X}},\bm{\mathsf{P}}\right)^T$. The electronic Hamiltonian is defined as, 
\begin{align}
   \mathrm{h}_{el}(\bm{z},\bm{X}) & = \frac{1}{2}\left[\bm{x}^T\cdot\bm{V}(\bm{X})\cdot\bm{x} + \bm{p}^T\cdot\bm{V}(\bm{X})\cdot\bm{p}  - \text{Tr}\left[\bm{V}(\bm{X})\right]\right]  \notag \\
    & + U\left(\bm{X}\right) . \label{eq:h_el}
\end{align}
The Wigner DHK correlation function, introduced in Paper I,\cite{Malpathak2024b} can be used with MMST mapping to yield,
\begin{align}
    C_{AB}^{\text{WDHK}}(t)  & = \frac{1}{\left(2\pi\hbar\right)^{2N}} \int d\bm{\bar{\mathsf{Z}}}_0 \int d\bm{\Delta \mathsf{Z}}_0\,\tilde{\rho}^{*}_{A}(\bm{\bar{\mathsf{Z}}}_0,\bm{\Delta \mathsf{Z}}_0)  \notag \\
    & \times \tilde{\mathcal{C}}_t(\bm{\bar{\mathsf{Z}}}_0,\bm{\Delta \mathsf{Z}}_0) \tilde{B}(\bm{\bar{\mathsf{Z}}}_t,\bm{\Delta \mathsf{Z}}_t) e^{i\tilde{S}_t(\bm{\bar{\mathsf{Z}}}_0,\bm{\Delta \mathsf{Z}}_0)/\hbar}. \label{eq:dhk-wig}
\end{align}
As defined in paper I,\cite{Malpathak2024b}, $\tilde{\rho}^{*}_{A}(\bm{\bar{\mathsf{Z}}}_0,\bm{\Delta \mathsf{Z}}_0)$ and $\tilde{B}(\bm{\bar{\mathsf{Z}}}_t,\bm{\Delta \mathsf{Z}}_t)$ are Wigner transforms smoothed by complex Gaussian functions,
\begin{align}
    \tilde{B}(\bm{\bar{\mathsf{Z}}}_t,\bm{\Delta \mathsf{Z}}_t) = \int d \bm{\mathsf{Z}}^{\prime} B_W(\bm{\mathsf{Z}}^{\prime}) g\left(\bm{\mathsf{Z}}^{\prime};\bm{\bar{\mathsf{Z}}}_t,\bm{\Delta \mathsf{Z}}_t\right). \label{eq:b_tilda}
\end{align}
$\tilde{\rho}^{*}_{A}(\bm{\bar{\mathsf{Z}}}_0,\bm{\Delta \mathsf{Z}}_0)$ is defined similarly, and the $*$ denotes complex conjugate. The complex Gaussian function is defined as,
\begin{align}
    g\left(\bm{z};\bm{\bar{\mathsf{Z}}},\bm{\Delta \mathsf{Z}}\right) & = \text{det}\left(\frac{\bm{\Gamma}}{\pi\hbar}\right)^{1/2} e^{-\frac{1}{\hbar}\left(\bm{\mathsf{Z}}-\bm{\bar{\mathsf{Z}}}\right)^{T}\cdot\bm{\Gamma}\cdot\left(\bm{\mathsf{Z}}-\bm{\bar{\mathsf{Z}}}\right)} \notag \\
    & \times e^{i\bm{\Delta \mathsf{Z}}^{T}\cdot\bm{J}^{T}\cdot\left(\bm{\mathsf{Z}}-\bm{\bar{\mathsf{Z}}}\right)/\hbar}, \label{eq:comp_gaus}
\end{align}
where $\bm{\Gamma}$ is a $2N \times 2N$ dimensional width matrix defined as, 
\begin{align}
    \bm{\Gamma} & = \left(\begin{array}{cc}
    \bm{\gamma}\hbar & \mathbb{0} \\
    \mathbb{0} & \frac{1}{\hbar}\bm{\gamma}^{-1} \\
    \end{array}\right), & \text{and} & &  \bm{J} &= \left(\begin{array}{rr}
     \mathbb{0} & \mathbb{1} \\
    -\mathbb{1} & \mathbb{0} \\
    \end{array}\right), \label{eq:j_and_gamma}
\end{align}
is the  $2N \times 2N$ dimensional symplectic matrix. In Eq.~\eqref{eq:dhk-wig}, $\tilde{\rho}^{*}_{A}(\bm{\bar{\mathsf{Z}}}_0,\bm{\Delta \mathsf{Z}}_0)$ represents $\hat{\rho}_{A}$ in the extended phase space of the mean ($\bm{\bar{\mathsf{Z}}}$)  and difference ($\bm{\Delta \mathsf{Z}}$) variables. These are related to the phase space variables of the forward and backward trajectories as: $\bm{\bar{\mathsf{Z}}} = \left(\bm{\mathsf{Z}}^{+} + \bm{\mathsf{Z}}^{-}\right)/2$ and $ \bm{\Delta \mathsf{Z}} = \bm{\mathsf{Z}}^{+} - \bm{\mathsf{Z}}^{-} $. Similarly, $\tilde{B}(\bm{\bar{\mathsf{Z}}}_t,\bm{\Delta \mathsf{Z}}_t)$ represents $\hat{B}$  in the extended phase space evaluated at time $t$ at the end points of the forward and backward trajectories $\bm{\mathsf{Z}}^{\pm}_t$ transformed into the mean and difference variables. The trajectories $\bm{\mathsf{Z}}^{\pm}_t$ follow the equations of motion, 
\begin{align}
    \dot{\bm{X}^{\pm}} & = \bm{m}^{-1}\cdot\bm{P}^{\pm}, \\
    \dot{\bm{P}^{\pm}} & = -\frac{\partial \bm{\mathrm{h}}_{el}(\bm{z}^{\pm},\bm{X}^{\pm})}{\partial \bm{X}^{\pm}},  \\
    \dot{\bm{z}}^{\pm} & = \bm{V}\left(\bm{X}^{\pm}\right)\cdot\bm{J}_e\cdot\bm{z}^{\pm} 
\end{align}
where $\bm{J}_e$ is a $2F \times 2F$ dimensional symplectic matrix in the electronic variable space. These equations conserve energy corresponding to the classical Hamiltonian $H_{MMST}(\bm{\mathsf{Z}}^{\pm})$. The prefactor $\tilde{\mathcal{C}}_t$ and the action $\tilde{S}_t$ are related to their respective counterparts in the conventional DHK expression as,
\begin{align}
    \tilde{\mathcal{C}}_t(\bm{\mathsf{Z}}^{+}_0,\bm{\mathsf{Z}}^{-}_0) & = \mathcal{C}_t\left(\bm{\mathsf{Z}}^{+}_0\right) \mathcal{C}^{*}_t\left(\bm{\mathsf{Z}}^{-}_0\right), \\ 
    \tilde{S}_t(\bm{\mathsf{Z}}^{+}_0,\bm{\mathsf{Z}}^{-}_0) & = -\bm{\bar{\mathsf{P}}}^{T}_t\cdot\bm{\Delta \mathsf{X}}_t + \bm{\bar{\mathsf{P}}}^{T}_0\cdot\bm{\Delta \mathsf{X}}_0 \notag \\ 
    & + S_t(\bm{\mathsf{Z}}^{+}_0) - S_t(\bm{\mathsf{Z}}^{-}_0), \label{eq:s_tilda}
\end{align}
where $S_t(\bm{\mathsf{Z}}^{\pm}_0)$ is the action of forward (backward) trajectory and 
the prefactor is the product of forward and backward HK prefactors individually defined 
as,
\begin{align}
    & \mathcal{C}_{t}\left(\bm{\mathsf{Z}}_{0}\right) \notag \\
    & = \text{det}\left|\frac{1}{2}\left[\bm{M}_{\mathsf{XX}}  + \bm{M}_{\mathsf{PP}}  - i\hbar\gamma\cdot\bm{M}_{\mathsf{XP}}+\frac{i}{\hbar}\gamma^{-{1}}\cdot\bm{M}_{\mathsf{PX}}\right]\right|^{1/2}.
    \label{eq:hk_pre}
\end{align}
For more details of the Wigner DHK framework, including the interpretation of the smoothed Wigner functions, refer to Paper I.\cite{Malpathak2024b} The equations of motion for the classical trajectories and their monodromy matrices under the MMST Hamiltonian, written in mean and difference variables, $(\bm{\bar{\mathsf{Z}}}_t,\bm{\Delta \mathsf{Z}}_t)$, will be used in subsequent sections and are provided in Appendix~\ref{ap:dhk_mmst}.

\subsection{nonadiabatic Hybrid Wigner Dynamics with Quantized Electronic System: NHWD(E)}
In this section we derive the NHWD(E) approximation where the nuclear dofs are treated as the classical bath and the electronic state dofs comprise the quantum system. 
This choice is motivated by the success MQC methods like Ehrenfest dynamics and surface hopping that are known to describe nonadiabatic problems with reasonable accuracy while treating nuclei classically. An important difference being that in the NHWD(E) framework we still capture important
nuclear quantum effects like zero-point energy and tunneling, only omitting nuclear quantum coherence effects. 

As we did when deriving the AHWD approximation for adiabatic processes, we derive NHWD(E) by performing a stationary phase approximation (SPA) over the nuclear variables. Details of this 
are provided in Appendix~\ref{app:nhwde}. The resulting NHWD(E) approximation to 
the correlation function is, 
\begin{align}
    C_{AB}^{\text{NHWD(E)}}(t) & = \frac{1}{\left(2\pi\hbar\right)^{2F + D}} \int d\bm{z}_{0}^{\pm} \int d\bm{\bar{Z}}_{0} \notag \\
    &
    \times 
    \left[\hat{\rho}_{A_N}\right]_W\left(\bm{\bar{Z}}_{0}\right)   \left[\hat{B}_N\right]_W\left(\bm{\bar{Z}}_{t}\right) \notag \\  
    & 
    \times \Tilde{{\rho}}_{A_e}^{*}(\bm{z}_{0}^{\pm}) \Tilde{B_e}(\bm{z}_{t}^{\pm}) \notag \\
    & \times \tilde{\mathcal{C}}_{t}^{\text{NHWD(E)}}\left(\bm{z}_{0}^{\pm},\bm{\bar{Z}}_{0}\right)e^{i\tilde{S}_t^{\text{NHWD(E)}}\left(\bm{z}_{0}^{\pm},\bm{\bar{Z}}_{0}\right)/\hbar} . \label{eq:cab_nhwde} 
\end{align}
In Eq.~\eqref{eq:cab_nhwde}, we have assumed that the operator $\hat{B}\equiv \hat{B}_e \otimes \hat{B}_N$ factorizes into nuclear and electronic parts and similarly for $\hat{\rho}_{A}$, where the subscripts $e$ and $N$ refer to operators corresponding to the electronic and nuclear dofs respectively. We note that this assumption is not required but is made here for simplicity. 
As a result of the SPA, the difference variables for the nuclear dofs are constrained to be zero, 
$\bm{\Delta Z}_{t} =0$ resulting in an LSC-like expression for the nuclei. Specifically, the nuclear operators are represented in mean variable phase space by their respective Wigner transforms, $\left[\hat{\rho}_{A_N}\right]_W\left(\bm{\bar{Z}}_{0}\right)$ at time zero and 
$ \left[\hat{B}_N\right]_W\left(\bm{\bar{Z}}_{t}\right)$ evaluated at time $t$ in the second line of Eq.~\eqref{eq:cab_nhwde}. 
The electronic dofs are described by a DHK-like correlation function where 
electronic operators are represented in the extended forward-backward phase space by their respective smoothed Wigner functions, $\Tilde{{\rho}}_{A_e}^{*}(\bm{z}_{0}^{\pm})$ and $\Tilde{B_e}(\bm{z}_{t}^{\pm})$, that appear in the third line of Eq.~\eqref{eq:cab_nhwde}. 
The structure of the trajectories is such that electronic dofs have separate forward and backward trajectories, $\bm{z}_{t}^{\pm}$, whereas, the nuclear variables are described by mean phase-space trajectories, $\bm{\bar{Z}_{t}}$. 
This aspect is highlighted in the equations of motion,
\begin{align}
    \dot{\bm{\bar{X}}} & = \bm{m}^{-1}\cdot\bm{\bar{P}}, \label{eq:nhwde_eom1}\\
    \dot{\bm{\bar{P}}} & = -\frac{1}{2}\left[\bm{\mathrm{h}}_{el}^{\prime}(\bm{z}^{+},\bm{\bar{X}}) + \bm{\mathrm{h}}_{el}^{\prime}(\bm{z}^{-},\bm{\bar{X}})\right], \\
    \dot{\bm{z}}^{\pm} & = \bm{V}\left(\bar{\bm{X}}\right)\cdot\bm{J}_e \cdot\bm{z}^{\pm} \label{eq:nhwde_eom4}
\end{align}
that conserve the Hamiltonian (not written in conjugate variables),
\begin{align}
    \bar{H}(\bm{{z}}^{\pm},\bm{\bar{Z}}) &= {\bm{\bar{P}}}^T\cdot\bm{m}^{-1}\cdot\bm{\bar{P}}  \notag \\
    & + \mathrm{h}_{el}(\bm{z}^+,\bm{\bar{X}})  + \mathrm{h}_{el}(\bm{z}^-,\bm{\bar{X}}),
\end{align}
and are used to compute the action and prefactor in line 4 of Eq.~\eqref{eq:cab_nhwde}
defined in the Appendix in Eq.~\eqref{eq:nhwde_act} and Eq.~\eqref{eq:nhwde_pref} respectively.

It is important to note that, as detailed in the derivation in Appendix~\ref{app:nhwde}, for the stationary phase conditions to hold, the electronic-nuclear (system-bath) and nuclear-electronic (bath-system) blocks of the monodromy matrix must be set to zero, as done in AHWD in Paper I. 
In order to ensure that the prefactor depends only on the electronic
dofs and can be computed using only the electronic-electronic block
of the monodromy matrix, we make a further approximation that 
$\mathrm{h}_{-}^{\prime\prime}(\bm{z}^{\pm},\bar{\bm{X}}) = 0$. 
For many widely used models of chemical systems like the spin-boson model,\cite{Leggett1987} the Frenkel exciton model,\cite{Frenkel1931} and the linear-vibronic coupling model,\cite{Kouppel1984} where the state-dependent part of the potential $\bm{V}^{\prime\prime}(\bm{X}) =0$, this condition is automatically satisfied.
The approximation ensures that the nuclear-nuclear block of the prefactor cancels exactly with the the hessian of the phase obtained when performing the SPA. 
Owing to the harmonic nature of the Hamiltonian for the electronic phase space variables, simplifications can be made to the electronic block of the monodromy matrix, as detailed in Appendix~\ref{app:nhwde}. The resulting prefactor is time-evolved according to the equation of motion, Eq.~\eqref{eq:nhwde_M_eom} that obey the symplecticity condition in Eq.~\eqref{eq:nhwde_M_symp}. 

As with AHWD, owing to the vanishing nuclear difference trajectory, 
we expect the overall phase to be smaller in magnitude than for 
the corresponding DHK expression. For electronic systems coupled to a large number of nuclear bath modes, we find a significant numerical advantage when compared to a full DHK calculation indicating that 
the mixed quantized framework used here mitigates the sign problem. 

Other approximate quantum dynamics methods for nonadiabatic problems that treat nuclear dofs classically while maintaining forward-backward electronic phase space trajectories, like Partially Linearized Density Matrix (PLDM)\cite{Huo2011,Huo2012} and the Forward-Backward Trajectory Solution (FBTS)\cite{Hsieh2012,Hsieh2013} also feature very similar equations of motion. In PLDM, the final term containing $\text{Tr}[\bm{V\left(\bar{X}\right)}]$ in $\mathrm{h}_{el}$, which pertains to the zero-point energy of the mapped electronic oscillators, is absent. FBTS is identical to PLDM, when the symmetrized version of the mapping Hamiltonian is used.\cite{Hsieh2012,Hsieh2013,Kelly2012} NHWD(E) differs from these methods in the presence of a SC prefactor, the complex action term, and the smoothed Wigner functions for operators 
$\hat{\rho}_{A_e}$ and $\hat{B}_e$. It is possible that with further simplifications to these expressions, especially with specific choices of $\bm{\Gamma}_e$, as done in PLDM, we may find further connections between these methods. Nonetheless, the similarity in the trajectory structure of NHWD(E) with PLDM suggests improved performance over LSC, and this is indeed the case as will be highlighted in model systems in Sec.~\ref{sec:results}. 

One of the key advantages of the HWD approach is its versatility in 
choosing which dofs to treat as the quantum system. We demonstrate
this by deriving a version where we quantize only the nuclear dofs
while treating the electronic dofs in the quantum-limit: the resulting NHWD(N) approximation is discussed in Appendix ~\ref{app:nhwdn}. 
A more natural and important system-bath partitioning is the vibronic system case where we choose to include electronic dofs {\it and} a handful of nuclear dofs in the quantum system and treat the remaining
nuclear modes as the classical bath. We discuss this case in detail in the following section.

\subsection{nonadiabatic Hyrbrid Wigner Dynamics with Quantized Vibronic System: NHWD(V)}

For many problems in condensed phase chemistry a few important vibrational dofs mediate nonadiabatic energy and population transfer, whereas a vast majority of environmental/bath modes only contribute indirectly through their coupling to the vibrational dofs of interest. In such problems, it is important to quantize the complete vibronic system \textemdash ~consisting of electronic dofs and the important vibrational dofs \textemdash ~while treating the bath modes classically to make the calculation computationally feasible. In this section we derive the NHWD(V) apporximation where we quantize a vibronic system. 

We partition the nuclear phase-space vector $\bm{Z}$, into the ``important" dofs included in the vibronic system $\bm{Z}_s$, and those included in the bath $\bm{Z}_b$. The complete vibronic system then consists of the electronic dofs and the important vibrational dofs, $\bm{\mathsf{Z}}^T_s \equiv \left(\bm{z},\bm{Z}_s\right)^T$. The full system-bath phase-space vector can then be written as $\bm{\mathsf{Z}}^T \equiv \left(\bm{\mathsf{Z}}_s,\bm{Z}_b\right)^T$.

To derive the NHWD(V) method, nuclear bath variables are treated using a stationary phase approximation, with details of the derivation provided in Appendix~\ref{app:nhwdv}.  The resulting NHWD(V) approximation to the quantum correlation function is, 
\begin{align}
    C_{AB}^{\text{NHWD(V)}}(t) & = \frac{1}{\left(2\pi\hbar\right)^{2N_s + N_b}} \int d\bm{\mathsf{Z}}_{0,s}^{\pm} \int d\bm{\bar{Z}}_{0,b} \notag\\
    & \times \left[\hat{\rho}_{A_b}\right]_W\left(\bm{\bar{Z}}_{0,b}\right)   \left[\hat{B}_b\right]_W\left(\bm{\bar{Z}}_{t,b}\right) \notag \\  
    &\times \Tilde{{\rho}}_{A_s}^{*}(\bm{\mathsf{Z}}_{0,s}^{\pm}) \Tilde{B_s}(\bm{\mathsf{Z}}_{t,s}^{\pm}) \notag \\
    & \times \tilde{\mathcal{C}}_{t}^{\text{NHWD(V)}}(\bm{\mathsf{Z}}_{0,s}^{\pm},\bm{\bar{Z}}_{0,b})e^{i\tilde{S}_t^{\text{NHWD(V)}}(\bm{\mathsf{Z}}_{0,s}^{\pm},\bm{\bar{Z}}_{0,b})/\hbar} . \label{eq:cab_nhwdv}
\end{align}
In Eq.~\eqref{eq:cab_nhwdv}, we have assumed that the operator $\hat{B}\equiv \hat{B}_s \otimes \hat{B}_b$ factorizes into nuclear and electronic parts and similarly for $\hat{\rho}_{A}$. 
This assumption is not required and has been made for the sake of simplicity. 
The structure of the correlation function corresponds to a DHK-like treatment for the vibronic system variables, and an LSC-like treatment for the nuclear bath variables. 
As a consequence of the stationary phase treatment of the bath, the difference variables for the nuclear bath dofs are constrained to be zero, $\bm{\Delta Z}_{t,b} =0$ and the nuclear bath operators are represented in mean variable phase space by their respective Wigner transforms, $\left[\hat{\rho}_{A_b}\right]_W\left(\bm{\bar{Z}}_{0,b}\right)$ at time zero and $ \left[\hat{B}_b\right]_W\left(\bm{\bar{Z}}_{t,b}\right)$ at time $t$. 
These appear in the second line of Eq.~\eqref{eq:cab_nhwdv}. 
Similar to a DHK calculation, the vibronic system operators are represented in the extended forward-backward phase space by their respective smoothed Wigner functions, $\Tilde{{\rho}}_{A_s}^{*}(\bm{\mathsf{Z}}_{0,s}^{\pm})$ and $\Tilde{B_s}(\bm{Z}_{t,s}^{\pm})$, that appear in the third line of Eq.~\eqref{eq:cab_nhwdv}. 

The structure of the trajectories is such that vibronic system 
variables are evolved through separate forward and backward trajectories, $\bm{\mathsf{Z}}_{t,s}^{\pm}$, whereas, the nuclear bath variables are described using mean phase space trajectories, $\bm{\bar{Z}_{t,b}}$. We emphasize here that the trajectories feel the \textit{\textbf{full}} potential along with the system bath coupling $U_{sb}(\bm{X})$ and $\bm{V}_{sb}(\bm{X})$ as highlighted in the equations of motion,	
\begin{align}
    \dot{\bm{{X}}}^{\pm}_s & = \bm{m}^{-1}_s\cdot\bm{{P}}^{\pm}_s, \\ \label{eq:nhwdv_eom1}
    \dot{\bm{{P}}}^{\pm}_s & = -\frac{\partial \bm{\mathrm{h}}_{el}(\bm{\mathsf{Z}}^{\pm}_s,\bm{\bar{X}_b})}{\partial \bm{X}^{\pm}_s }, \\
    \dot{\bm{\bar{X}}}_b & = \bm{m}^{-1}_b\cdot\bm{\bar{P}}_b, \\
    \dot{\bm{\bar{P}}}_b & = -\frac{1}{2}\left[\frac{\partial \bm{\mathrm{h}}_{el}(\bm{\mathsf{Z}}^{+}_s,\bm{\bar{X}_b})}{\partial \bm{\bar{X}}_b} + \frac{\partial \bm{\mathrm{h}}_{el}(\bm{\mathsf{Z}}^{-}_s,\bm{\bar{X}_b})}{\partial \bm{\bar{X}}_b}\right], \\
    \dot{\bm{z}}^{\pm} & = \bm{V}(\bm{X}^{\pm}_s,\bm{\bar{X}}_b)\cdot\bm{J}_e\cdot\bm{z}^{\pm}, \label{eq:nhwdv_eom2}
\end{align}
that conserve the Hamiltonian (not written in conjugate variables),
\begin{align}
    \bar{H}(\bm{\mathsf{Z}}^{\pm}_s,\bm{\bar{Z}_b}) & = \frac{1}{2}{\bm{P}^+_s}^T\cdot\bm{m}^{-1}_s\cdot\bm{P}^+_s + \frac{1}{2}{\bm{P}^-}^T_s\cdot\bm{m}^{-1}_s\cdot\bm{P}^-_s  \notag \\ 
    & +  {\bm{\bar{P}}}^T_b\cdot\bm{m}^{-1}_b\cdot\bm{\bar{P}}_b \notag \\
    &  + \mathrm{h}_{el}(\bm{\mathsf{Z}}^+_s,\bm{\bar{X}}_b) + \mathrm{h}_{el}(\bm{\mathsf{Z}}^-_s,\bm{\bar{X}}_b). \label{eq:nhwdv_ham}
\end{align}
Analogous to the NHWD(E), the off-diagonal system-bath blocks of the monodromy matrix are assumed to vanish to satisfy the stationary phase conditions arising from the SPA for the bath dofs. 
Effectively, \textit{\textbf{only}} for the propagation of 
the monodromy matrices, we assume the system-bath couplings in the potential vanish, $U_{sb}(\bm{X}) = \bm{V}_{sb}(\bm{X}) = 0 $.
The state-independent and state-dependent system bath couplings are defined in Eq.~\eqref{eq:pot-split}. Enforcing this yields a prefactor expression that can be separated into system and bath prefactors.
As with NHWD(E), we further facilitate the cancellation of the bath prefactor with the Hessian from the SPA by assuming $\frac{\partial^2 \mathrm{h}_{-}}{\partial \bm{\bar{X}}_b^2} = 0$, implying $\bm{\Delta M}_{bb}(t)=0$. For standard models like the electron transfer model\cite{Topaler1996}, the spin-boson model,\cite{Leggett1987}, the linear-vibronic coupling model,\cite{Kouppel1984} and  models of exciton-polariton chemistry,\cite{Chowdhury2021} $\bm{V}_b^{\prime\prime}(\bm{X}_b) = 0$, and this condition is automatically satisfied.  With this approximation, following similar arguments from the AHWD case in paper I,\cite{Malpathak2024b} the bath block of the monodromy matrix is not required in the calculation. We thus only need to propagate the system monodromy matrix, which follows the equations of motion, Eq.~\eqref{eq:nhwdv_mono_eom} and the symplecticity condition, Eq.~\eqref{eq:nhwdv_symp}. It is used to calculate the NHWD(V) prefactor, $\tilde{\mathcal{C}}_{t}^{\text{NHWD(V)}}(\bm{\mathsf{Z}}_{0,s}^{\pm},\bm{\bar{Z}}_{0,b})$, which is defined in Eq.~\eqref{eq:nhwdv_pref}. 

In cases where the vibronic system consists of a select few important nuclear dofs and the bath consists of a large number of dofs representing an environment, the fact that the prefactor only requires the system monodromy matrix elements is a considerable reduction in the cost of the calculation, as compared to a full DHK calculation. Moreover, noting that the form of the prefactor, Eq.~\eqref{eq:nhwdv_pref}, is the same as that of the DHK prefactor for the forward-backward system trajectories, well established approximations to it are available to accomplish a further reduction in numerical cost.\cite{Gelabert2000d,Liberto2016b} 
The NHWD(V) action, $\tilde{S}_t^{\text{NHWD(V)}}(\bm{\mathsf{Z}}_{0,s}^{\pm},\bm{\bar{Z}}_{0,b})$, is defined in Eq.~\eqref{eq:nhwde_act},
and can be expected to be smaller in magnitude than the full DHK action in Eq.~\eqref{eq:s_tilda}, leading to a less oscillatory integral in Eq.~\eqref{eq:cab_nhwdv} when compared to a full DHK calculation. However, the action is expected to be not as small as the NHWD(E) case, making the NHWD(V) correlation function more expensive to converge as may be expected from expanding the system size, in general. Lastly, we note that a similar approximate scheme capable of quantizing a vibronic system has been introduced by \citeauthor{Provazza2019}.\cite{Provazza2019} However, the method involves approximations to the prefactor and the connection with DHK theory are not apparent.

\section{Model Systems and Simulation Details}\label{sec:simulation}

\begin{figure}
    \centering
    \includegraphics[width=0.5\textwidth]{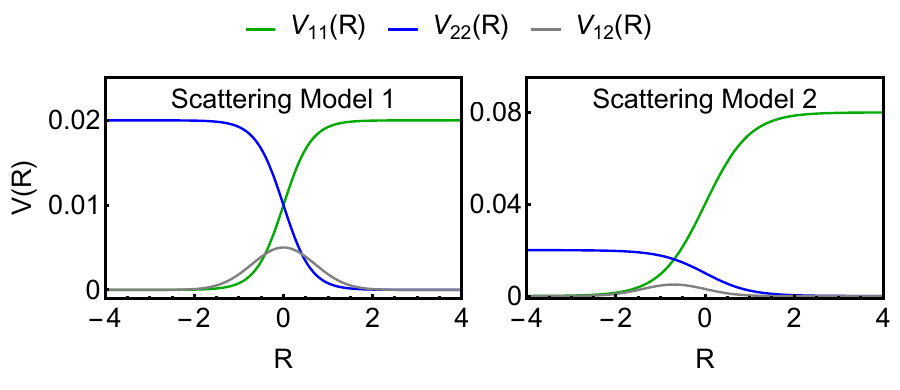}
    \caption{Plots showing the matrix elements of the diabatic potential energy matrix for the scattering models 1 (left) and 2 (right). The legend highlights the colors corresponding to the diabatic states $V_{11}(R)$ and $V_{22}(R)$, and the coupling $V_{12}(R)$. } 
    \label{fig:tully_pots}
\end{figure}

We test NHWD(E) and NHWD(V) on standard benchmark models for nonadiabatic dynamic methods. We start with scattering models 1 and 2, that feature an avoided crossing,~\cite{Ananth2007} with one nuclear dof $(R)$  and two electronic states. The diabatic potentials are given by, 
\begin{align}
    V_{11}(R) & = V_1 \left(1+\text{tanh}(\alpha R)\right), \\
    V_{22}(R) & = V_2 \left(1-\text{tanh}(\alpha R)\right), \\
    V_{12}(R) & = ae^{-b(R+f)^2},
\end{align}
with the parameters for the two model given in Table.~\ref{tbl:tully_params}. Plots for the two models are shown in Fig.~\ref{fig:tully_pots}. For both models the nuclear dof is initialized in a coherent state centered at $R_i = -5$, and initial momentum $P_i$ directed towards the avoided crossing. Two cases are considered \textemdash ~the low energy case with kinetic energy 0.03 a.u. corresponding to $P_i = 10.9$ a.u., and the high energy case with kinetic energy 0.1 a.u., corresponding to $P_i = 19.9$ a.u. The width of the nuclear coherent state is chosen to be $\gamma_i = 0.25$ a.u., and the nuclear mass is taken to be the mass of a proton, $m = 1980$ a.u. The system is assumed to start in state 1, with the electronic part of the initial density taken to be a projection onto state 1, yielding a density matrix, $\hat{\rho} = \ketbra{\bm{Z}_i;\gamma_i}{\bm{Z}_i;\gamma_i}\otimes\ketbra{1}{1}$. For both models, we calculate time dependent expectation values of the Pauli spin matrices $\langle {\bm{\sigma}}_i(t) \rangle$ for $i \in \{x,y,z\}$.  Exact quantum dynamic calculations performed using a Colbert-Miller discrete variable representation (DVR) grid are used as a benchmark.\cite{Colbert1992a}
We use the scattering models to test the performance of the NHWD(E) approximation. 

\begin{table}
    \centering
    \caption{Parameters for scattering models 1 and 2 in atomic units.}
  \label{tbl:tully_params}
 
  \begin{tabular}{|c|c|c|c|c|c|c|}
  \hline
    Model   &  $V_1$    &  $V_2$ & $\alpha$ & $a$ & $b$ & $f$   \\
    \hline \hline 

    Scattering model 1 & 0.01 & 0.01 & 1.6 & 0.005 & 1.0 & 0 \\ \hline
    Scattering model 2 & 0.04 & 0.01 & 1.0 & 0.005 & 1.0 & 0.7 \\ \hline \hline
  \end{tabular}
  
\end{table}

\begin{figure*}
    \centering
    \includegraphics[width=0.95\textwidth]{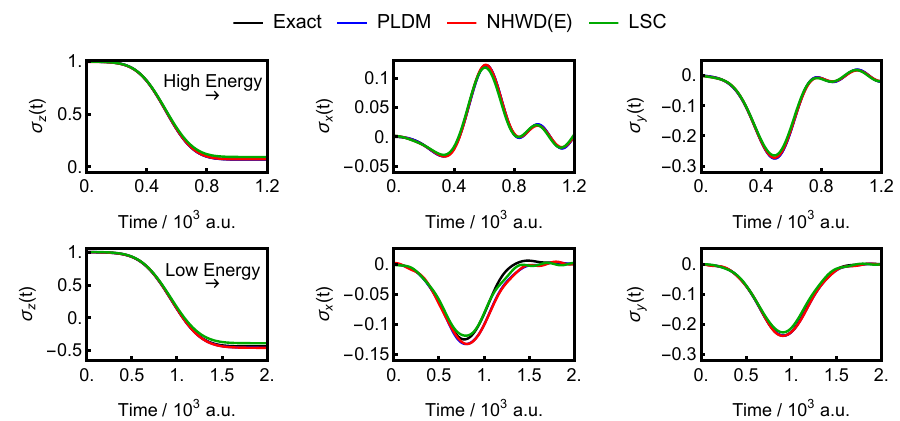}
    \caption{Expectation values as a function of time are plotted for the Pauli spin matrices $\bm{\sigma}_z$ (left column), $\bm{\sigma}_x$ (middle column) and $\bm{\sigma}_y$ (right column) for scattering  model 1. Top and bottom rows corresponds to high and low initial kinetic energy cases respectively. Exact quantum dynamical calculations performed on a DVR grid\cite{Colbert1992a} are compared with NHWD(E) (red), PLDM (blue) and LSC (green). NHWD(E) and PLDM overlap with each other in all panels.} 
    \label{fig:tully_m1}
\end{figure*}

We also study the spin-boson model \cite{Leggett1987}, which is a prototypical model for nonadiabatic energy transfer in a condensed phase environment. The Hamiltonian is,
\begin{align}
    \bm{H}(\bm{X},\bm{P}) &= \sum_{j=1}^D \left[\frac{P_j^2}{2m_j} + \frac{1}{2}m_j\omega_j^2 X_j^2 \right] {\bm{\mathbb{1}}} \notag \\
    & + \left(\epsilon + \sum_{j=1}^D c_j X_j\right){\bm{\sigma}}_z + \Delta {\bm{\sigma}}_x,  
\end{align}
with $(X_j,P_j)$, $j \in \left[1,D\right]$ corresponding to the nuclear positions and momenta of the $j^{th}$ mode respectively, $\epsilon$ is the energy bias parameter between the two electronic states, and $\Delta$ is the constant coupling between them. The nuclear dofs consist of $D$ harmonic oscillators with  $m_j =1$, frequencies  $\omega_j$, and coupling constants $c_j$. All parameters are in atomic units. The spectral density of the nuclear modes determines their frequencies. Here we employ an Ohmic spectral density with exponential cut-off, \cite{Leggett1987} 
\begin{align}
     J_{bath}(\omega) = \frac{\pi\xi}{2}\omega e^{-\omega/\omega_c}
\end{align}
where $\omega_c$ is the characteristic frequency and $\xi$ is the Kondo parameter that determines the strength of the coupling to the electronic variables. The spectral density is discretized into $D$ modes as, 
\begin{align}
    J_{bath}(\omega) = \frac{\pi}{2}\sum_{j=1}^D \frac{c_j^2}{m_j\omega_j}\delta(\omega- \omega_j),
\end{align}
using the discretization procedure from Craig and Manolopoulos.~\cite{Craig2005} Convergence was achieved with $D=50$ for all models considered here. We investigate 6 models, named (a)-(f), that cover a wide range of physical regimes: high and low temperature, symmetric and asymmetric electronic states, and weak versus strong coupling to nuclear modes.\cite{Cotton2016a,Runeson2019,Mannouch2020} Parameters for these models are provided in Table~\ref{tbl:spin-boson_params}. The initial density is obtained by assuming a thermal distribution of nuclear modes with zero system-bath coupling and with a single electronic state $\ket{1}$ occupied,
\begin{align}
	\hat{\rho} = \frac{e^{-\beta H_{nuc}}}{\text{Tr}\left[e^{-\beta H_{nuc}}\right]} \otimes \ketbra{1}{1}
\end{align}
where $ H_{nuc}(\bm{\hat{Z}})\equiv \sum_{j=1}^{D}\frac{\hat{P}_j^2}{2m_j} + \frac{1}{2}m_j\omega_j^2 \hat{X}_j^2$. For all models, we calculate time dependent expectation values of the Pauli spin matrices $\langle {\bm{\sigma}}_i(t) \rangle$ for $i \in \{x,y,z\}$. As a benchmark,we perform exact quantum dynamic calculations. For models (a)-(d) we use SMatPI \cite{Makri2020,Makri2020b,Makri2021} with QUAPI,\cite{Makri1992} and for models (e)-(f) we use SMatPI with Blipsum, \cite{Makri2014} as implemented in the PathSum code package. \cite{Makri1995,Makri2014,Makri2020,Kundu2023}

\begin{table}
    \centering
    \caption{Parameters for spin-boson models (a)-(f). 
 $\Delta = 1$ a.u. in all models. All parameters are in atomic units. }
  \label{tbl:spin-boson_params}
 
  \begin{tabular}{|c|c|c|c|c|c|c|c|}
  \hline
    Model   &  $\epsilon$ &  $\xi$ & $\beta$ & $\omega_c$ & $\Omega$ & $\omega_1 $ & $\omega_2$ \\
    \hline \hline 

    a & 0 & 0.09 & 0.1 & 2.5 & 2 & 1.996 & -\\ \hline
    b & 0 & 0.09 & 5   & 2.5 & 2 &  1.996 &  - 
    \\ \hline
    c & 1 & 0.1  & 5   & 2.5  & 2$\sqrt{2}$ & 2.771 & -\\ \hline
    d & 1 & 0.1  & 0.25& 1.0 & 2$\sqrt{2}$ & 2.659 & - 
    \\ \hline
    e & 0 & 2    & 1   & 1.0  &  2 & 2.040 & 0.994 \\ \hline
    f & 5 & 4    & 0.1 & 2.0  & 2$\sqrt{26}$ & 5.318 & 9.210\\ \hline
     \hline
  \end{tabular}
  
\end{table}

We note that although it is common to refer to the nuclear dofs as the thermal bath, we intentionally refrain from doing so and refer to them as the nuclear modes/dofs to avoid confusion with the different system-bath partitionings in NHWD(E) and NHWD(V). To test the effect of quantizing the electronic dofs, we use the NHWD(E) method, where all nuclear modes are treated as the classical bath. With the NHWD(V) method, we explore the effect of quantizing select nuclear modes on the electronic populations and coherences. Here, these select nuclear modes and the electronic variables constitute the quantized vibronic system and the rest of the nuclear modes constitute the classical bath. In the vibronic system, we choose 1-2 nuclear dofs with frequencies $\omega_j$ such that integral multiples of $\omega_j$ are nearly resonant with the Rabi frequency $(\Omega)$ of the electronic sub-system. Table~\ref{tbl:spin-boson_params} lists the Rabi frequency and frequencies of the quantized nuclear modes in NHWD(V) calculations ($\omega_1$ and $\omega_2$) for all the models. 

As is standard in the MMST mapping,\cite{Church2018} the electronic part of the initial density matrix is mapped onto singly excited harmonic oscillator states, $\ketbra{1}{1} \to \ketbra{1_1,0_2}{1_1,0_2}$. For NHWD(E), initial phase space points for the forward and backward trajectories of the electronic system, $\bm{z}^{\pm}_0$, are then sampled from the absolute magnitude of the coherent state matrix element, $\left|\braket{\bm{z}^{+}_0}{1_1,0_2}\braket{1_1,0_2}{\bm{z}^{-}_0}\right|$, and those for the mean trajectory of the nuclear bath, $\bm{\bar{Z}}_0$, are sampled from the Wigner transform of the initial nuclear coherent state, $\left[\ketbra{\bm{Z}_{i};\gamma_i}{\bm{Z}_{i};\gamma_i}\right]_W(\bar{Z}_0)$.

\begin{figure*}
    \centering
    \includegraphics[width=0.95\textwidth]{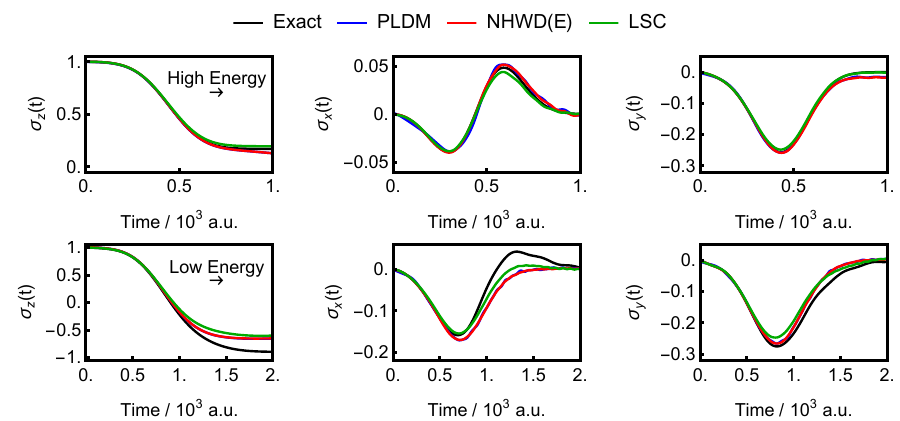}
    \caption{Expectation values as a function of time are plotted for the Pauli spin matrices $\bm{\sigma}_z$ (left column), $\bm{\sigma}_x$ (middle column) and $\bm{\sigma}_y$ (right column) for scattering  model 2. Top and bottom rows corresponds to high and low initial kinetic energy cases respectively. Exact quantum dynamical calculations are compared with NHWD(E) (red), PLDM (blue) and LSC (green). NHWD(E) and PLDM overlap with each other in all panels.} 
    \label{fig:tully_m2}
\end{figure*}

For the spin-boson models, for both NHWD(E) and NHWD(V), the initial phase space points for the forward and backward trajectories of the electronic system, $\bm{z}^{\pm}_0$, are sampled from the absolute magnitude of the coherent state matrix element, $\left|\braket{\bm{z}^{+}_0}{1_1,0_2}\braket{1_1,0_2}{\bm{z}^{-}_0}\right|$. For NHWD(E) all nuclear modes are treated as the classical bath, and thus the initial phase space points for their mean trajectories, $\bm{\bar{Z}}_{0,b}$, are sampled from the Wigner transform of the nuclear Boltzmann operator, $\left[e^{-\beta H_{nuc}(\hat{\bm{Z}}_b)}\right]_W(\bm{\bar{Z}}_{0,b})$. As mentioned earlier, for NHWD(V), the nuclear modes are partitioned into system and bath nuclear modes. The initial phase space points for the forward and backward trajectories of the system nuclear modes, $\bm{Z}^{\pm}_{0,s}$, are sampled from the absolute magnitude of the coherent state matrix element, 
$\left|\mel{\bm{Z}^{+}_{0,s}}{e^{-\beta H_{sys}(\hat{\bm{Z}}_s)}}{\bm{Z}^{-}_{0,s}}\right|$, whereas those for the mean trajectories of the classical bath nuclear modes, $\bm{\bar{Z}}_{0,b}$, are sampled from the Wigner transform of the uncoupled bath Hamiltonian Boltzmann distribution, $\left[e^{-\beta H_{bath}(\hat{\bm{Z}}_b)}\right]_W(\bm{\bar{Z}}_{0,b})$. Note that here the Hamiltonian corresponding to the full set of uncoupled nuclear modes has been divided into system and bath parts as
$H_{nuc}(\bm{Z})  =  H_{sys}(\bm{Z}_s) + H_{bath}(\bm{Z}_b) $. Finally, we also note that for all the cases where initial phase space points are sampled from the absolute magnitude of a coherent state matrix element, we have used the connection between the smoothed Wigner transform of an operator and its coherent state matrix element from Paper I.\cite{Malpathak2024b} Sampling from the absolute magnitude of either is completely equivalent.

For all the models considered, unless otherwise stated, we use the symmetrized version of the mapping Hamiltonian,\cite{Kelly2012} which corresponds to partitioning the state dependent diabatic matrix such that Tr$[\bm{V}(\bm{X})]=0$. Classical trajectories are propagated using the symplectic MInt algorithm.\cite{Church2018} For the scattering models, a timestep of 1.5 a.u. and 0.5 a.u. is used for the high and low energy cases respectively. For all the spin boson models,  a timestep of 0.01 a.u. is used. Trajectories breaking energy conservation, as indicated using $\left| 1 - E(t)/E(0)\right| \geq E_{tol}$, or breaking the symplecticity for the system monodromy matrices within a tolerance of $S_{tol}$ are discarded. For the scattering models, $E_{tol} = S_{tol} = 10^{-4}$, whereas for the spin-boson models, $E_{tol} = S_{tol} = 10^{-3}$. For both scattering models and spin-boson models (a),(d)-(f), $<1\%$ of total trajectories are discarded, whereas for the low-temperature spin-boson models, (b) and (c), $<2\%$ of total trajectories are discarded. For the scattering models, $10^6$ and $10^7$ trajectories are used to yield converged NHWD(E) results for models 1 and 2 respectively. Table~\ref{tbl:spin-boson_ntraj} presents the number of trajectories used in the spin-boson models. The large number trajectories required is due to the calculation of the $\langle\bm{\sigma}_x\rangle$ and $\langle\bm{\sigma}_y\rangle$, which are more expensive than $\langle\bm{\sigma}_z\rangle$ calculations. In general NHWD(E) and NHWD(V) calculations are more expensive than LSC calculations, but as discussed in the upcoming section, also yield improved results. 

\begin{table}
    \centering
    \caption{Number of trajectories used for calculations for spin-boson models (a)-(f). NHWD(V)-1 and NHWD(V)-2 correspond to NHWD(V) calculations with 1 and 2 nuclear modes included in the vibronic system, as described in the text.}
  \label{tbl:spin-boson_ntraj}
 
  \begin{tabular}{|c|c|c|c|c|}
  \hline
    Model   &  LSC &  NHWD(E) & NHWD(V)-1 &  NHWD(V)-2  \\
    \hline \hline 

    a & $10^7$ & $10^7$ & $10^7$ & - \\ \hline
    b & $10^6$ & $10^6$ & $10^6$ & - \\ \hline
    c & $10^6$ & $10^6$ & $10^6$ & - \\ \hline
    d & $10^6$ & $10^6$ & $10^6$ & - \\ \hline
    e & $10^6$ & $10^6$ & $10^6$ & $10^7$ \\ \hline
    f & $10^7$ & $10^7$ & $10^7$ & $10^7$  \\ \hline
     \hline
  \end{tabular}
  
\end{table}

For both models, to calculate the time-dependent expectation value of the  Pauli spin matrices $\langle {\bm{\sigma}}_i(t) \rangle$ for $i \in \{x,y,z\}$, we use ${\bm{\sigma}}_z = \ketbra{1}{1}-\ketbra{2}{2}$, ${\bm{\sigma}}_x = 2\text{ Re}(\ketbra{1}{2})$ and ${\bm{\sigma}}_y = 2\text{ Im}(\ketbra{1}{2})$. These are then mapped onto harmonic oscillator states as prescribed in the MMST mapping, $\hat{B}_e \equiv \ketbra{i}{j} \to \ketbra{1_i,0_{k \neq i}}{1_j,0_{k \neq j}}$.\cite{Church2018} Not  that $\hat{B}_N = \hat{\mathbb{1}}$. The results presented here are all normalized such that Tr$_e[\hat{\rho}_e(t)] = 1$ at all times.

\section{Results and Discussion} \label{sec:results}
\begin{figure*}
    \centering
    \includegraphics[width=0.95\textwidth]{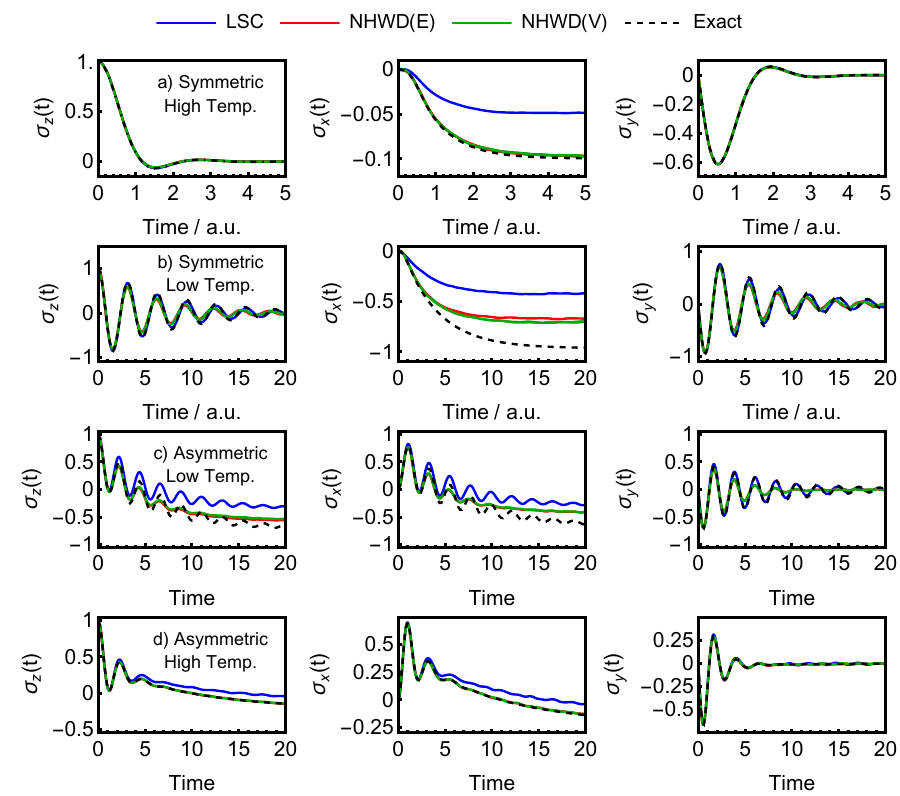}
    \caption{Expectation values as a function of time are plotted for the Pauli spin matrices $\bm{\sigma}_z$ (left column), $\bm{\sigma}_x$ (middle column) and $\bm{\sigma}_y$ (right column) for spin-boson models (a)-(d). Calculations performed using LSC (blue), NHWD(E) (red), and NHWD(V) (green) are compared against exact results. NHWD(E) and NHWD(V) results overlap whenever they are not clearly distinguishable.}
    \label{fig:spin_boson}
\end{figure*}
\subsection{Scattering Models}
For the scattering models, we track the expectation values of the Pauli spin matrices as a function of time using three approximate dynamic methods \textemdash~NHWD(E), PLDM, and LSC,~\textemdash and we 
compare the results against numerically exact calculations performed on a DVR grid.\cite{Colbert1992a} In Fig.~\ref{fig:tully_m1} we plot the expectation values for the three Pauli matrices for Tully Model 1 for both high and low initial kinetic energy cases. In both cases the initial kinetic energy is sufficient to cross the potential barrier and minimal reflection is expected. LSC calculations are already very accurate for this model, and PLDM and NHWD(E) are seen to marginally improve LSC results for both $\bm{\sigma}_z$ and  $\bm{\sigma}_y$. Interestingly, PLDM and NHWD(E) results are identical in all panels, and require similar number of trajectories for convergence.

In Fig.~\ref{fig:tully_m2} we plot results for scattering model 2. For this model, the high kinetic energy case has sufficient high energy to allow transmission on state 1, whereas for the low energy case, significant reflection is expected. Owing to the presence of both transmission and reflection channels for the energies considered, this model is a more stringent test for 
the methods in question than model 1. 
LSC is seen to be fairly accurate for the high-energy case, with both PLDM and NHWD(E) offering marginal improvements. For the low energy case, LSC fails to capture the full extent of population transfer, and the magnitude of the coherences is also smaller in LSC calculations when compared against the exact result. Both PLDM and NHWD(E) predict more population transfer than LSC, which is in slightly better agreement with the exact result. Similarly, the amplitude of $\bm{\sigma}_y$ is also predicted better by PLDM and NHWD(E) than LSC. Again, PLDM and NHWD(E) provide identical results.

\begin{figure*}
    \centering
    \includegraphics[width=0.98\textwidth]{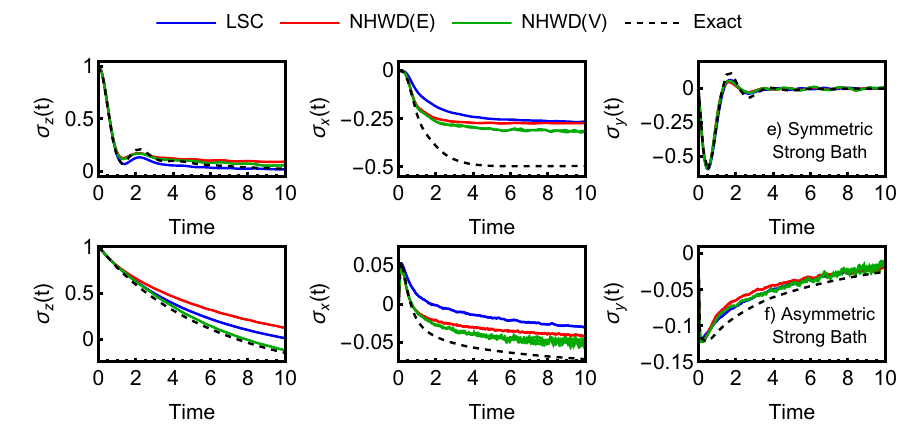}
    \caption{Expectation values as a function of time are plotted for the Pauli spin matrices $\bm{\sigma}_z$ (left column), $\bm{\sigma}_x$ (middle column) and $\bm{\sigma}_y$ (right column) for spin-boson models (e) \&(f). Calculations performed using LSC (blue), NHWD(E) (red), and NHWD(V) (green) are compared against exact results.}
    \label{fig:spin_boson2}
\end{figure*}

The calculations with the scattering models are meant to serve as a simple test of the NHWD(E) method and it is seen to yield comparable results to LSC. Interestingly, for both models and both kinetic energy cases considered, NHWD(E) results are identical to those obtained using PLDM. In Appendix~\ref{app:comp_hams}, we present calculations performed using the standard MMST Hamiltonian, and see that NHWD(E) and PLDM yield distinct results. Our calculations with the symmetrized Hamiltonian seem to suggest deeper connections between NHWD(E) and PLDM when a symmetrized Hamiltonian is used. Investigations into these similarities, and the discrepancies between LSC and NHWD(E) results, especially for $\bm{\sigma}_x$ will be the subject of future studies.

\subsection{Spin-Boson Models}

In Fig.~\ref{fig:spin_boson}, we plot the expectation values of the Pauli spin matrices calculated using LSC, NHWD(E) and NHWD(V). We note that for all spin-boson models, (a)-(f), results using NHWD(E) are identical to PLDM results (not shown). For NHWD(V) calculations shown in Fig.~\ref{fig:spin_boson}, one nuclear dof resonant with the Rabi frequency of the electronic sub-system ($\omega_1$ in Table ~\ref{tbl:spin-boson_params}) is quantized.

We first consider the symmetric models with weak coupling to the bath, models (a) and (b). For these models, LSC is very accurate for $\langle {\bm{\sigma}}_z \rangle$ and $\langle {\bm{\sigma}}_y \rangle$. Both  NHWD(E)/(V) preserve this accuracy and also improve upon LSC for $\langle {\bm{\sigma}}_x \rangle$ which underestimates the long-time plateau value for both these models. For the high temperature case, model (a), NHWD(E)/(V) both yield essentially exact results for all three Pauli operators. For the low temperate case, model (b), NHWD(E)/(V) predict slightly damped oscillations for $\langle {\bm{\sigma}}_z \rangle$ and $\langle {\bm{\sigma}}_y \rangle$. It has been observed that employing the spin-mapping framework can remedy this issue.\cite{Mannouch2020} Moreover, both NHWD(E) and NHWD(V) are unable to capture the correct long-time plateau of $\langle {\bm{\sigma}}_x  \rangle$. It has been shown that employing an identity trick with LSC,\cite{Saller2019,Gao2020} or using spin-mapping methods like spin-LSC or spin-PLDM can improve the long-time plateau value.\cite{Mannouch2020} For the high temperature asymmetric models with weak coupling, model (d), NHWD(E)/(V) improve on LSC to  yield exact results. In the low temperature asymmetric model (c), NHWD(E)/(V) improve upon LSC, but do not match exact results, especially at longer times, and notably, predict oscillations that are damped compared to exact results. However, similar to model (b), these can be improved using the identity-trick or the use of the spin-mapping framework.\cite{Saller2019,Gao2020,Mannouch2020}

For these models with weak-coupling to the bath, NHWD(E), which yields identical results to PLDM, provides significant improvements to LSC. Moreover, in cases where good agreement with exact results is not observed, it has already been shown employing an identity trick with MMST mapping, or switching to the spin mapping-framework can yield significant improvements. These observations suggest that the shortcomings of the NHWD(E) method stem from the MMST framework, and not from the approximations in the semiclassical HWD method. Lastly, quantizing a bath mode resonant with the Rabi frequency of the electronic sub-system within the NHWD(V) framework does not yield any significant improvement in accuracy.

Models (e) and (f) have stronger coupling to the nuclear bath compared to the models considered so far. In these models NHWD(E), which again is identical to PLDM (not shown), does not reliably provide an improvement to LSC. This is reasonable considering that the assumption of vanishing electronic-nuclear monodromy matrix blocks made in NHWD(E) gets worse for strong coupling. However, if one nuclear dof resonant with the Rabi frequency of the electronic sub-system ($\omega_1$ in Table ~\ref{tbl:spin-boson_params}) is also quantized within the NHWD(V) framework, significant improvements over LSC are observed. Specifically, for both models (e) and (f), NHWD(V) significantly improves $\langle {\bm{\sigma}}_z \rangle$ results, while also marginally improving $\langle {\bm{\sigma}}_x \rangle$. 

\begin{figure}
    \centering
    \includegraphics[width=0.47\textwidth]{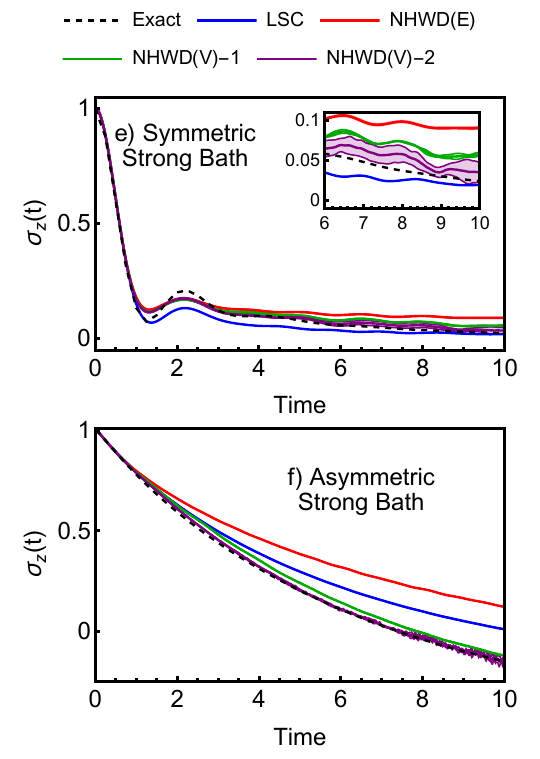}
    \caption{Expectation values as a function of time are plotted for the Pauli spin matrix $\bm{\sigma}_z$ for spin-boson models (e) \& (f). Calculations performed using LSC (blue), NHWD(E) (red), NHWD(V) with one quantized nuclear dof, denoted as NHWD(V)-1 (green), and NHWD(V) with two quantized nuclear dofs, denoted as NHWD(V)-2 (purple), are compared against exact results.}
    \label{fig:spin_boson3}
\end{figure}

So far, we have only quantized one nuclear dof resonant with the Rabi frequency of the electronic sub-system in NHWD(V) simulations. Here we inspect the effect of quantizing a second nuclear dof resonant with the Rabi frequency, with their 
 frequency presented in 
Table~\ref{tbl:spin-boson_params}. For weak-coupling models (a)-(d), we see no improvement over quantizing just one mode (not shown). In Fig.~\ref{fig:spin_boson3}, we plot the expectation values of $\langle {\bm{\sigma}}_z \rangle$ for the strong-coupling models (e) and (f). For model (e), quantizing a second nuclear dof improves the long-time accuracy of NHWD(V), as seen in the inset. For model (f), quantizing the a second nuclear dof brings NHWD(V) in agreement with the exact QUAPI results for the time studied here. No improvements are seen in $\langle {\bm{\sigma}}_x \rangle$ and $\langle {\bm{\sigma}}_y \rangle$ on quantizing a second nuclear dof (not shown).

\section{Conclusions} \label{sec:conclusion}

In this article, we have introduce two novel SC dynamic methods for nonadiabatic problems, NHWD(E) and NHWD(V), in conjunction with the MMST Hamiltonian. Both methods have been derived from the full DHK correlation by treating the bath dofs using a stationary phase approximation. Subsequent approximations that facilitate the reduction in the dimensions of the SC prefactor for both methods have also been outlined. Interestingly, the trajectory structure that emerges for NHWD(E) is very similar to the well known PLDM and FBTS methods. This result is not surprising, but provides insights into the approximations inherent in these approaches. As evidenced by the derivation of the NHWD(E) method, the quantized treatment of the electronic dofs in both PLDM and FBTS inherently assumes that the electronic-nuclear coupling is ignored in the semiclassical prefactor for these methods. Moreover, the observation that both PLDM and NHWD(E) yield identical results for all models studies in this article when a symmetrized MMST Hamiltonian is used, suggests that deeper connections between these methods might exist and need to be investigated. 

The NHWD(V) method provides an avenue to study the role of quantized nuclear dofs in nonadiabatic problems. The hybrid nature of the approach, which treats a majority of the nuclear dofs as a classical bath, facilitates numerical convergence by significantly mitigating the sign problem. It is important to note that for the spin-boson models studied here, the Quantum-Classical Liouville Equation (QCLE) provides an exact description of the dynamics.\cite{MacKernan2002,Shi2004,Kapral2015} This suggests that if the electronic dofs and their coupling to the nuclear dofs is treated exactly within the QCLE, a classical treatment of the nuclear dofs is sufficient to provide an exact description of the spin boson model. However, in the context of the SC methods described in this article, `quantizing' the system dofs refers to a full forward-backward SC treatment, not an exact quantum-mechanical treatment. Furthermore, with the NHWD framework, the system-bath interactions are also approximated in the SC prefactor.
Thus, the fact that `quantizing' just the electronic dofs in the NHWD(E) method does not yield exact results for the spin-boson model is not surprising. Moreover, in conjunction with the results presented for the strong-coupling spin boson models, this analysis suggests that although quantizing the nuclear modes is not required with an exact treatment of the electronic dofs, a SC treatment \textit{does} benefit from nuclear quantization.

Preliminary investigations using the spin-boson models for both NHWD(E) and NHWD(V) presented in this article highlight that both methods show great promise for further investigations. In particular, NHWD(E) does not require the nuclear Hessian matrix, and can serve as a potential improvement over \textit{ab-initio} LSC calculations.\cite{Miyazaki2023} NHWD(V) can be used to investigate the role of nuclear quantum effects in models of electron-transfer and the role of quantizing the photonic dofs in the exciton-polariton chemistry.\cite{Chowdhury2021} Futhermore, given its versatility in choosing the vibronic system, NHWD(V) is perfectly suited to the study of ultrafast relaxation dynamics involving conical intersections in conjunction with vibronic coupling models for molecular systems.\cite{Koppel1993,Raab1999,Izmaylov2011,Liu2020,Gomez2024} 
Investigations into the accuracy and computational benefits offered by approximations to the SC prefactor,\cite{Gelabert2000d,Liberto2016b} along with phase filtering techniques to facilitate numerical convergence\cite{Makri1987c,Makri1988b,Church2018,Malpathak2022} can potentially bring down the cost of these SC methods and pave the way for their widespread use in simulations of nonadiabatic chemistry.

\begin{acknowledgments}
The authors thank Prof. Gregory Ezra for many stimulating discussions on quantum dynamics in phase space. S.M. acknowledges funding from the Cornell University Department of Chemistry and Chemical Biology.
\end{acknowledgments}
\appendix

\section{Nonadiabatic DHK in Mean-Difference Variables} \label{ap:dhk_mmst}

As discussed in Paper I,\cite{Malpathak2024b} a DHK correlation function generally involves calculation of a forward backward trajectory pair, $\mathsf{Z}_t^{\pm}$. However, to derive variants of NHWD, we need to rewrite the expression in terms of mean and difference variables, $(\bm{\bar{\mathsf{Z}}}_t,\bm{\Delta \mathsf{Z}}_t)$. Following paper I,\cite{Malpathak2024b}, the sum Hamiltonian is defined similarly as in the Born-Oppenheimer case, but this time using the MMST Hamiltonian, Eq.~\eqref{eq:h_mmst},
\begin{align}
    \bar{H}_{MMST}(\bm{\mathsf{Z}}^{\pm}) & = \frac{1}{2}{\bm{P}^+}^T\cdot\bm{m}^{-1}\cdot\bm{P}^+ + \frac{1}{2}{\bm{P}^-}^T\cdot\bm{m}^{-1}\cdot\bm{P}^- \notag \\ 
    & + \mathrm{h}_{el}(\bm{z}^+,\bm{X}^+) + \mathrm{h}_{el}(\bm{z}^-,\bm{X}^-). \label{eq:h-sum}
\end{align}
It propagates a pair of forward-backward trajectories, $\mathsf{Z}_t^{\pm}$, and can also be converted to mean and difference variables to propagate the mean and difference trajectories, $(\bm{\bar{\mathsf{Z}}}_t,\bm{\Delta \mathsf{Z}}_t)$. Here it is presented in forward-backward variables $\bm{\mathsf{Z}}^{\pm}$ for the sake of compactness. To obtain the equations of motion for mean and difference variables, we need to obtain their respective conjugate momenta. This can be done following a similar procedure as that followed in Paper I, to obtain the equations of motion for the nuclear variables, 
\begin{align}
    \dot{\bm{\bar{X}}} & = \bm{m}^{-1}\cdot\bm{\bar{P}}, \\ \label{eq:dhk-eom1}
    \bm{\Delta}\dot{\bm{X}} & =\bm{m}^{-1}\cdot\bm{\Delta P}, \\
    \dot{\bm{\bar{P}}} & = -\frac{1}{2} \bm{\mathrm{h}}_{+}^{\prime}(\bm{z}^{\pm},\bm{X}^{\pm}), \\
    \bm{\Delta}\dot{\bm{P}} & = - \bm{\mathrm{h}}_{-}^{\prime}(\bm{z}^{\pm},\bm{X}^{\pm}),
\end{align}
where $\mathrm{h}_{\pm} = \mathrm{h}_{el}(\bm{z}^+,\bm{X}^+) \pm \mathrm{h}_{el}(\bm{z}^-,\bm{X}^-)$ and $^{\prime}$ denotes derivative with respect to the nuclear position.  Similar to the Born-Oppenheimer case, the nuclear mean and difference trajectories feel a mean and difference of the electronic Hamiltonian. For the electronic variables the equations of motion are,
\begin{align}
     \dot{\bm{\bar{z}}} & = \frac{1}{2}\bm{V}_{+}\cdot\bm{J}_e\cdot\bm{\bar{z}} + \frac{1}{4}\bm{V}_{-}\cdot\bm{J}_e\cdot\bm{\Delta z}, \\
     \bm{\Delta}\dot{\bm{z}} & = \frac{1}{2} \bm{V}_{+}\cdot\bm{J}_e\cdot\bm{\Delta z} + \bm{ V}_{-}\cdot\bm{J}_e\cdot\bm{\bar{z}}, \label{eq:dhk-eom2}
\end{align}
where $\bm{V}_{\pm} = \bm{V}\left(\bm{X}^{+}\right) \pm \bm{V}\left(\bm{X}^{-}\right)$ and $\bm{J}_e$ is a $2F \times 2F$ dimensional symplectic matrix. It is important to note here that $\bm{\bar{\mathsf{X}}}$ and $\bm{\bar{\mathsf{P}}}$ are not conjugate variables, and the same is true for the difference positions and momenta,\cite{Malpathak2024b} but the equations of motion, Eqs.~\eqref{eq:dhk-eom1} -~\eqref{eq:dhk-eom2} conserve the Hamiltonian, Eq.~\eqref{eq:h-sum} written in these non-conjugate variables, $\bar{H}_{MMST}(\bm{\bar{\mathsf{Z}}},\bm{\Delta \mathsf{Z}}).$
The forward-backward action is,
\begin{align}
{S}_t(\bm{\mathsf{Z}}^{+}_0) - {S}_t(\bm{\mathsf{Z}}^{-}_0) & = \int_0^{t} d\tau \, \left[ \bm{\bar{P}}_{\tau}\cdot\bm{m}^{-1}\cdot\bm{\Delta P}_{\tau} \right. \notag \\
& \left. + \dot{\bm{x}}^{+}_{\tau}\bm{p}^{+}_{\tau} -  \dot{\bm{x}}^{-}_{\tau}\bm{p}^{-}_{\tau} - \mathrm{h}_{-}(\bm{z}^{\pm}_{\tau},\bm{X}^{\pm}_{\tau})  \right].
\end{align}
Written in a mean-difference variable basis, $\bm{\mathsf{Z}}_{md}^T \equiv \left(\bm{\bar{\mathsf{Z}}},\bm{\Delta \mathsf{Z}}\right)^T$, the monodromy matrix can be written as,\cite{Malpathak2024b} 
\begin{align}
    \mathcal{M}_{md} = \left(\begin{array}{cc}
    \bm{\bar{M}} & \bm{\Delta M} \\
    4 \bm{\Delta M}  &  \bm{\bar{M}} \\
    \end{array}\right),
\end{align}
where $\bm{\bar{M}} = \bm{M}_{\bm{\bar{\mathsf{Z}}}\bm{\bar{\mathsf{Z}}}} = \bm{M}_{\Delta\bm{\mathsf{Z}}\Delta\bm{\mathsf{Z}}}$ and $\bm{\Delta M} = \bm{M}_{\bm{\bar{\mathsf{Z}}}\Delta\bm{\mathsf{Z}}} = \frac{1}{4}\bm{M}_{\Delta\bm{\mathsf{Z}}\bm{\bar{\mathsf{Z}}}}$.  The structure of the monodromy matrix is similar to the Born-Oppenheimer case. The equations of motion also have a similar structure, 
\begin{align}
    \dot{\mathcal{M}}_{md} = \left(\begin{array}{cc}
    \mathbb{A} &  \frac{1}{4}\mathbb{B} \\
    \mathbb{B}  &  \mathbb{A} \\
    \end{array}\right)\cdot
    \mathcal{M}_{md}, \label{eq:eom_mono_dhk}
\end{align}
with,
\begin{align}
    \mathbb{A} & =\left(\begin{array}{cccc}
    \mathbb{0} &  \frac{\partial\dot{\bm{\bar{x}}}}{\partial \bm{\bar{X}}} & \frac{1}{2}\bm{V}_{+} & \mathbb{0}  \\
    \mathbb{0}  &  \mathbb{0}  &  \mathbb{0} & \bm{m}^{-1} \\
    - \frac{1}{2}\bm{V}_{+} & 
 - \frac{\partial\dot{\bm{\bar{p}}}}{\partial \bm{\bar{X}}} & \mathbb{0}  &  \mathbb{0} \\
  - \frac{\partial\dot{\bm{\bar{p}}}}{\partial \bm{\bar{X}}}^T & -\frac{1}{2}\mathrm{h}_+^{\prime\prime} & -\frac{\partial\dot{\bm{\bar{x}}}}{\partial \bm{\bar{X}}}^T & \mathbb{0}
    \end{array}\right),   
\end{align}
and,
\begin{align}
    \mathbb{B} & = 4\left(\begin{array}{cccc}
    \mathbb{0} &  \frac{\partial\dot{\bm{\bar{x}}}}{\partial \bm{\Delta X}} & \frac{1}{4}\bm{V}_{-} & \mathbb{0}  \\
    \mathbb{0}  &  \mathbb{0}  &  \mathbb{0} &  \mathbb{0} \\
    - \frac{1}{4}\bm{V}_{-} & 
 - \frac{\partial\dot{\bm{\bar{p}}}}{\partial \bm{\Delta X}} & \mathbb{0}  &  \mathbb{0} \\
   \frac{1}{4} \frac{\partial\bm{\Delta}\dot{\bm{{p}}}}{\partial \bm{\bar{X}}}^T & -\frac{1}{4}\mathrm{h}_-^{\prime\prime} & -\frac{1}{4}\frac{\partial\bm{\Delta}\dot{\bm{{x}}}}{\partial \bm{\bar{X}}}^T & \mathbb{0}
    \end{array}\right),
\end{align}
with intial conditions $\bm{\bar{M}}(0)=\mathbb{1}$ and $\bm{\Delta M}(0) = 0$. The monodromy matrix follows the symplecticity conditions,
\begin{align}
    \bm{\bar{M}}^T\cdot\bm{J}\cdot\bm{\bar{M}} + 4\bm{\Delta M}^T\cdot\bm{J}\cdot\bm{\Delta M} & = \bm{J}, \label{eq:sym_mb_dm1}
\end{align}
and, 
\begin{align}
    \bm{\bar{M}}^T\cdot\bm{J}\cdot\bm{\Delta M} = \left(\bm{\bar{M}}^T\cdot\bm{J}\cdot\bm{\Delta M}\right)^T. \label{eq:sym_mb_dm2} 
\end{align}
The following identities relating some of the blocks of the matrices $\mathbb{A}$ \& $\mathbb{B}$ will be used in deriving variants of NHWD,
\begin{align}
\frac{\partial\dot{\bm{\bar{x}}}}{\partial \bm{\bar{X}}} &= \frac{\partial\bm{\Delta}\dot{\bm{{x}}}}{\partial \bm{\Delta {X}}} = \frac{1}{2}\bm{V}_{+}^{\prime}\cdot\bm{\bar{p}} + \frac{1}{4}\bm{V}_{-}^{\prime}\cdot\bm{\Delta p},   \label{eq:der_id1}\\
\frac{\partial\dot{\bm{\bar{p}}}}{\partial \bm{\bar{X}}} &= \frac{\partial\bm{\Delta}\dot{\bm{{p}}}}{\partial \bm{\Delta {X}}} = - \left[\frac{1}{2}\bm{V}_{+}^{\prime}\cdot\bm{\bar{x}} + \frac{1}{4}\bm{V}_{-}^{\prime}\cdot\bm{\Delta x}\right],  \\ 
\frac{\partial\dot{\bm{\bar{x}}}}{\partial \bm{\Delta X}} &= \frac{1}{4}\frac{\partial\bm{\Delta}\dot{\bm{{x}}}}{\partial \bm{\bar{X}}} = \frac{1}{4}\bm{V}_{-}^{\prime}\cdot\bm{\bar{p}} + \frac{1}{8}\bm{V}_{+}^{\prime}\cdot\bm{\Delta p}, \\
\frac{\partial\dot{\bm{\bar{p}}}}{\partial \bm{\Delta X}} &= \frac{1}{4}\frac{\partial\bm{\Delta}\dot{\bm{{p}}}}{\partial \bm{\bar{X}}} = -\left[ \frac{1}{4}\bm{V}_{-}^{\prime}\cdot\bm{\bar{x}} + \frac{1}{8}\bm{V}_{+}^{\prime}\cdot\bm{\Delta x} \right]. \label{eq:der_id4}
\end{align}

\section{Deriving NHWD(E).} \label{app:nhwde}
Here we derive the NHWD(E) approximation from nonadiabatic Wigner-DHK correlation function,
\begin{align}
    C_{AB}^{W-DHK}(t)  & = \frac{1}{\left(2\pi\hbar\right)^{2N}} \int d\bm{\bar{\mathsf{Z}}}_0 \int d\bm{\Delta \mathsf{Z}}_0 \int d\bm{\mathsf{Z}}\int d\bm{\mathsf{Z}}^{\prime}\,\left[\hat{\rho}_{A}\right]_W(\bm{\mathsf{Z}}) \notag \\
    & g^{*}\left(\bm{\mathsf{Z}};\bm{\bar{\mathsf{Z}}}_0,\bm{\Delta \mathsf{Z}}_0\right) B_W(\bm{\mathsf{Z}}^{\prime})g\left(\bm{\mathsf{Z}}^{\prime};\bm{\bar{\mathsf{Z}}}_t,\bm{\Delta \mathsf{Z}}_t\right) \notag \\ & \times \tilde{\mathcal{C}}_t(\bm{\bar{\mathsf{Z}}}_0,\bm{\Delta \mathsf{Z}}_0) e^{i\tilde{S}_t(\bm{\bar{\mathsf{Z}}}_0,\bm{\Delta \mathsf{Z}}_0)/\hbar} , \label{eq:cab_dhk_mmst}
\end{align}
and performing the integrals over the nuclear variables $\bm{\Xi}^T = \left(\bm{Z},\bm{Z}^{\prime},\bm{\Delta Z}_0\right)^T$ by the stationary phase approximation (SPA) while keeping the integrals over the electronic variables, $\bm{\xi}^T = \left(\bm{z},\bm{z}^{\prime},\bm{\Delta z}_0\right)^T$ exact. The integrals over all the mean variables $\bm{\bar{\mathsf{Z}}}_0$ are performed exactly. The phase $\phi\left(\bm{\mathsf{\xi}}\equiv\left(\bm{\xi},\bm{\Xi}\right)\right)$ in the SPA, as in AHWD, is identified as the action $\tilde{S}_t(\bm{\bar{\mathsf{Z}}}_0,\bm{\Delta \mathsf{Z}}_0)$ along with exponential contributions from $g^{*}\left(\bm{\mathsf{Z}};\bm{\bar{\mathsf{Z}}}_0,\bm{\Delta \mathsf{Z}}_0\right)$ and  $g\left(\bm{\mathsf{Z}}^{\prime};\bm{\bar{\mathsf{Z}}}_t,\bm{\Delta \mathsf{Z}}_t\right)$, 
\begin{align}
    \phi(\bm{\mathsf{\xi}}) & = i\left(\bm{\mathsf{Z}}-\bm{\bar{\mathsf{Z}}}_0\right)^{T}\cdot\bm{\Gamma}\cdot\left(\bm{\mathsf{Z}}-\bm{\bar{\mathsf{Z}}}_0\right) - \bm{\Delta \mathsf{Z}}_0^{T}\cdot\bm{J}^{T}\cdot\left(\bm{\mathsf{Z}}-\bm{\bar{\mathsf{Z}}}_0\right) \notag \\
    & + i\left(\bm{\mathsf{Z}}^{\prime}-\bm{\bar{\mathsf{Z}}}_t\right)^{T}\cdot\bm{\Gamma}\cdot\left(\bm{\mathsf{Z}}^{\prime}-\bm{\bar{\mathsf{Z}}}_0\right) \notag \\
    &  + \bm{\Delta \mathsf{Z}}_t^{T}\cdot\bm{J}^{T}\cdot\left(\bm{\mathsf{Z}}^{\prime}-\bm{\bar{\mathsf{Z}}}_t\right)  + \tilde{S}_t(\bm{\bar{\mathsf{Z}}}_0,\bm{\Delta \mathsf{Z}}_0). \label{eq:phi}
\end{align}
The stationary phase equations emerge as,
\begin{align}
     \frac{\partial \phi}{\partial \bm{Z}} & = 2i\bm{\Gamma}_N\cdot\left(\bm{Z}-\bm{\bar{Z}}_0\right) + \bm{J}_N^T\cdot\bm{\Delta Z}_{0} = 0, \label{eq:HWDE_spc1} \\ 
      \frac{\partial \phi}{\partial \bm{Z}^{\prime}} & = 2i\bm{\Gamma}_N\cdot\left(\bm{Z}^{\prime}-\bm{\bar{Z}}_t\right) + \bm{J}_N^T\cdot\bm{\Delta Z}_{t} = 0,   \label{eq:HWDE_spc2} \\
      \frac{\partial \phi}{\partial \bm{\Delta Z}_{0}}  & = \bm{J}_N\cdot\left(\bm{Z}-\bm{\bar{Z}}_0\right) + \left[\bm{\bar{M}}^T\cdot\bm{J}^T\cdot\left(\bm{\mathsf{Z}}^{\prime}-\bm{\bar{\mathsf{Z}}}_t\right)\right]_N \notag \\
     & -2i\left[\bm{\Delta M}^T\cdot\bm{\Gamma}\cdot\left(\bm{\mathsf{Z}}^{\prime}-\bm{\bar{\mathsf{Z}}}_t\right)\right]_N = 0,  \label{eq:HWDE_spc3}
\end{align} 
where the subscript $N$ denotes the nuclear components of the vector, or nuclear-nuclear block of a matrix.
Analogous to AHWD, the stationary phase conditions emerge to be 
\begin{align}
    \bm{Z} &= \bm{\bar{Z}}_{0}, & \bm{Z}^{\prime} &= \bm{\bar{Z}}_{t},  & \text{and} &  & \bm{\Delta Z}_{t} &= 0. \label{eq:NHWDe__sp_conditions} 
\end{align}
However, these conditions are not sufficient to satisfy all three SP equations. To satify Eq.~\eqref{eq:HWDE_spc3}, we must make a further approximation that the system-bath and bath-system blocks of the monodromy matrices vanish. Here these correspond to the electronic-nuclear (eN) and nuclear-electronic (Ne) blocks of the monodromy matrices. Following similar analysis as in Paper I, this can be achieved if the following blocks of matrices $\mathbb{A}$ and $\mathbb{B}$ vanish:
\begin{align}
    \mathbb{A}_{eN} & = \left(\begin{array}{rr}
    \frac{\partial\dot{\bm{\bar{x}}}}{\partial \bm{\bar{X}}} &  \mathbb{0} \\
    - \frac{\partial\dot{\bm{\bar{p}}}}{\partial \bm{\bar{X}}} &  \mathbb{0} \\
    \end{array}\right), \notag \\ 
    \mathbb{A}_{Ne} & = \left(\begin{array}{cc}
      \mathbb{0} &  \mathbb{0} \\
    - \frac{\partial\dot{\bm{\bar{p}}}}{\partial \bm{\bar{X}}}^T & - \frac{\partial\dot{\bm{\bar{x}}}}{\partial \bm{\bar{X}}}^T  \\
    \end{array}\right), \notag \\
     \mathbb{B}_{eN} & = 4\left(\begin{array}{rr}
    \frac{\partial\dot{\bm{\bar{x}}}}{\partial \bm{\Delta {X}}} &  \mathbb{0} \\
    - \frac{\partial\dot{\bm{\bar{p}}}}{\partial \bm{\Delta {X}}} &  \mathbb{0} \\
    \end{array}\right), \notag \\
    \mathbb{B}_{Ne} & = \left(\begin{array}{cc}
      \mathbb{0} &  \mathbb{0} \\
     \frac{\partial\bm{\Delta}\dot{\bm{{p}}}}{\partial \bm{\bar{X}}}^T & - \frac{\partial\bm{\Delta}\dot{\bm{{x}}}}{\partial \bm{\bar{X}}}^T  \\
    \end{array}\right). \label{eq:ab_nad}
\end{align}
Setting these blocks to zero amounts to assuming that the rate of change of the electronic phase space variables is independent of the nuclear positions, or equivalently that the force felt by the quantized electronic states is independent of the position of the classical bath. This assumption is similar in spirit to the vanishing system-bath coupling assumption in the AHWD case. It is important to note that this assumption is\textit{ \textbf{only}} made in the propagation of the monodromy matrices, \textbf{\textit{not}} in the trajectories themselves. For the special case that $\bm{V}(\bm{X})$ is a constant matrix (electronic states uncoupled from the nuclei), this condition is automatically satisfied. With these assumptions in place, the stationary phase conditions Eqs.~\eqref{eq:HWDE_spc1}-~\eqref{eq:HWDE_spc3} are satisfied. As a consequence, the classical nuclear dofs have only a mean trajectory, $\bar{\bm{Z}}_t$, with the difference trajectory being constrained to be $\bm{\Delta Z}_t =0$. The equations of motion are presented in the main text in Eqs.~\eqref{eq:nhwde_eom1}-\eqref{eq:nhwde_eom4}.

In the case of AHWD in Paper I, the assumption of vanishing system-bath coupling that led to the system-bath blocks of the monodromy matrices to vanish, also yielded the simplification that $\bm{\Delta M}_{bb} = 0$.\cite{Malpathak2024b} This simplification allowed the cancellation of the bath part of the prefactor with the Hessian of the phase appearing from the SPA. In the case of NHWD(E), the assumptions that lead to vanishing off-diagonal blocks of the monodromy matrices, that is setting all the matrices in Eq.~\eqref{eq:ab_nad} to zero, separate the prefactor into electornic and nuclear prefactors, but \textbf{do not} lead to $\bm{\Delta M}_{bb} \equiv \bm{\Delta M}_{NN} = 0$. Thus, in order to simplify the prefactor, we need to set $\bm{\Delta M}_{NN}(t) = 0$, by making a further approximation, $\mathrm{h}_{-}^{\prime\prime} = 0$. Since $\mathrm{h}_{-}^{\prime\prime} \propto \frac{\partial \dot{\bm{\bar{P}}}_t}{\partial\bm{\Delta X}_t}$, this amounts to assuming that the force on the mean nuclear trajectory is independent of the nuclear difference variable. For systems where $\bm{V}^{\prime\prime}(\bm{X}) =0$, which is the case in many models mentioned in the main text, this condition is already satisfied. With this assumption, $\bm{\Delta M}_{NN}(t) = 0$, and following a similar reasoning as in AHWD the bath part of the prefactor cancels the Hessian of the phase. 

Furthermore, given that the MMST Hamiltonian is harmonic in the electronic phase space variables, the monodromy matrix for the electronic variables follows the equation of motion, 
\begin{align}
    \dot{\bm{\bar{M}}}_{ee} & = \left(\begin{array}{cc}
    \mathbb{0} & \bm{V}(\bm{\bar{X}}) \\
    - \bm{V}(\bm{\bar{X}})  & \mathbb{0} \\
    \end{array}\right)\cdot\bm{\bar{M}}_{ee} \label{eq:nhwde_M_eom}
\end{align}
with initial condition $\bm{\bar{M}}_{ee}(0) = \mathbb{1}$, and the difference monodromy matrix simplifies to $\bm{\Delta M}_{ee}(t)=0$. The mean monodromy matrix satisfies the symplecticity condition, 
\begin{align}
 \bm{\bar{M}}_{ee}^T\cdot\bm{J}_{e}\cdot\bm{\bar{M}}_{ee}  = \bm{J}_{e}. \label{eq:nhwde_M_symp}
\end{align} 
The prefactor simplifies to Eq.~\eqref{eq:nhwde_pref}, and only requires the electronic-electronic monodromy matrix. Taken together the NHWD(E) approximation to a correlation function can be written as 
\begin{align}
    C_{AB}^{\text{NHWD(E)}}(t) & = \frac{1}{\left(2\pi\hbar\right)^{2F + D}} \int d\bm{z}_{0}^{\pm} \int d\bm{\bar{Z}}_{0} \, \Tilde{{\rho}}_{A_e}^{*}(\bm{z}_{0}^{\pm}) \Tilde{B_e}(\bm{z}_{t}^{\pm}) \notag \\
    & \times \left[\hat{\rho}_{A_N}\right]_W\left(\bm{\bar{Z}}_{0}\right)   \left[\hat{B}_N\right]_W\left(\bm{\bar{Z}}_{t}\right) \notag \\  & \times \tilde{\mathcal{C}}_{t}^{\text{NHWD(E)}}\left(\bm{z}_{0}^{\pm},\bm{\bar{Z}}_{0}\right)e^{i\tilde{S}_t^{\text{NHWD(E)}}\left(\bm{z}_{0}^{\pm},\bm{\bar{Z}}_{0}\right)/\hbar} .
\end{align}
Details on the phase space functions of the operators and trajectory structure are provided in the main text. The NHWD(E) action is,
\begin{align}
   & \tilde{S}_t^{\text{NHWD(E)}}(\bm{{z}}^{\pm}_0,\bm{\bar{Z}}_0)  \notag \\
    & \equiv -\bm{\bar{p}}^{T}_t\cdot\bm{\Delta x}_t + \bm{\bar{p}}^{T}_0\cdot\bm{\Delta x}_0 
     + {S}_t(\bm{{z}}^{+}_0,\bm{\bar{Z}}_0) - {S}_t(\bm{{z}}^{-}_0,\bm{\bar{Z}}_0) \notag \\
     & = -\bm{\bar{p}}^{T}_t\cdot\bm{\Delta x}_t + \bm{\bar{p}}^{T}_0\cdot\bm{\Delta x}_0  \notag \\
     & + \int_0^{t} d\tau \, \left[\dot{\bm{x}}^{+}_{\tau}\bm{p}^{+}_{\tau} -  \dot{\bm{x}}^{-}_{\tau}\bm{p}^{-}_{\tau} 
     - \mathrm{h}_{-}(\bm{z}^{\pm}_{\tau},\bar{\bm{X}}_{\tau})  \right], \label{eq:nhwde_act}
\end{align}
where $\mathrm{h}_{-}(\bm{z}^{\pm},\bar{\bm{X}}) = \mathrm{h}_{el}(\bm{z}^+,\bar{\bm{X}}) - \mathrm{h}_{el}(\bm{z}^-,\bar{\bm{X}})$. The NHWD(E) prefactor is, 
\begin{align}    
   & \tilde{\mathcal{C}}_{t}^{NHWD(E)}\left(\bm{z}_{0}^{\pm},\bm{\bar{Z}}_{0}\right) = \text{det}\left(2\bm{\Gamma}_e\right)^{-1/2} \notag \\
   & \times \text{det}\left[\bm{\Gamma}_e\cdot\bm{\bar{M}}_{ee}  -\bm{J}_e\cdot\bm{\bar{M}}_{ee}\cdot\bm{J}_e\cdot\bm{\Gamma}_e  \right]^{1/2}. \label{eq:nhwde_pref}
\end{align}
Only the electronic-electronic monodromy matrix, $\bm{\bar{M}}_{ee}$, is required, which follows the equation of motion given by in Eq.~\eqref{eq:nhwde_M_eom} and the symplecticity condition, Eq.~\eqref{eq:nhwde_M_symp}.

\section{Nonadiabatic Hybrid Wigner Dynamics with Quantized Nuclei and Classical Electronic States: NHWD(N)} \label{app:nhwdn}
It has been shown previously that quantizing the nuclei can accurately predict the branching of the nuclear wavepacket after encountering an avoided crossing\cite{Church2018}. Here, we derive a version of HWD with quantized nuclei and classical electronic states. We refer to this method as NHWD(N) with the first N referring to a nonadiabatic problem and the (N) referring to the quantized nature of the nuclear dofs.

We start from the full DHK-Wigner correlation function, Eq.~\eqref{eq:cab_dhk_mmst} and perform the integrals over the electronic dofs $\bm{\xi} = \left(\bm{z},\bm{z}^{\prime},\bm{\Delta z}_0\right)^T$ by SPA while keeping the integrals over the nuclear dofs $\bm{\Xi}^T = \left(\bm{Z},\bm{Z}^{\prime},\bm{\Delta Z}_0\right)^T$ exact. The integrals over all the mean variables $\bm{\bar{\mathsf{Z}}}_0$ are performed exactly, and the phase in the SPA is the same as in Eq.~\eqref{eq:phi}. As can be expected by now, the the stationary phase conditions emerge as,
\begin{align}
    \bm{z} &= \bm{\bar{z}}_{0}, & \bm{z}^{\prime} &= \bm{\bar{z}}_{t},  & \text{and} &  & \bm{\Delta z}_{t} &= 0. \label{eq:HWD_nuc_sp_conditions} 
\end{align}
As in the other cases, these conditions are not sufficient to satisfy all three SP equations. We must make a further approximation that the system-bath and bath-system blocks of the monodromy matrices vanish to satisfy all three SP equations. Here these blocks correspond to the nuclear-electronic (Ne) and electronic-nuclear (eN)  blocks of the monodromy matrices. Following analysis similar to that in the derivation of AHWD in Paper I,\cite{Malpathak2024b} this corresponds to setting the off-diagonal blocks of $\mathbb{A}$ and $\mathbb{B}$ in the system-bath basis, given in Eq.~\eqref{eq:ab_nad} to zero. Using Eqs.~\eqref{eq:der_id1}-\eqref{eq:der_id4} and the equations of motion, Eq.~\eqref{eq:nucf_HWDn}, this can be interpreted as ignoring the dependence of the nuclear mean and difference force on the electronic phase space variables \textbf{only} in the monodromy matrix propagation. Again, this is similar in spirit to assuming that the nuclear-electronic (system-bath) coupling vanishes for propagating monodromy matrices.

With these assumptions the stationary phase equations are satisfied and the trajectories follow the equations of motion,
\begin{align}
    \dot{\bm{{X}}}^{\pm} & = \bm{m}^{-1}\cdot\bm{{P}}^{\pm}, \\
    \dot{\bm{{P}}}^{\pm} & = -\bm{\mathrm{h}}_{el}^{\prime}(\bm{\bar{z}},\bm{X}^{\pm}), \label{eq:nucf_HWDn}\\
    \dot{\bm{\bar{z}}} & = \frac{1}{2}\bm{V}_+\cdot\bm{J}_e\cdot\bm{\bar{z}}  .
\end{align}
where $\bm{V}_{+} = \bm{V}\left(\bm{X}^{+}\right) + \bm{V}\left(\bm{X}^{-}\right)$ and $\bm{J}_e$ is a $2F \times 2F$ dimensional symplectic matrix.
The nuclear forward-backward trajectories both feel the effect of the mean electronic phase space variables, and in turn also influence its motion. The trajectories conserve the Hamiltonian (not written in conjugate variables), 
\begin{align}
    \bar{H}(\bm{Z}^{\pm},\bm{\bar{z}}) & = \frac{1}{2}{\bm{P}^+}^T\cdot\bm{m}^{-1}\cdot\bm{P}^+ + \frac{1}{2}{\bm{P}^-}^T\cdot\bm{m}^{-1}\cdot\bm{P}^- \notag \\ 
    & + \mathrm{h}_{el}(\bm{\bar{z}},\bm{X}^+) + \mathrm{h}_{el}(\bm{\bar{z}},\bm{X}^-),
\end{align}

The assumption of the vanishing off-diagonal monodromy matrix blocks separates the prefactor into nuclear and electronic prefactors. But like the NHWD(E) case, a further approximation is necessary so that the bath prefactor, which in this case is the electronic prefactor, cancels the Hessian from the SPA. As before, for this to be true, we require, $\bm{\Delta M}_{bb}(t)\equiv\bm{\Delta M}_{ee}(t) =0$ which can be achieved by setting,  $\bm{V}_{-} = 0$. This is a very peculiar condition which is only true for constant diabatic potential energy matrices. We interpret the appearance of this peculiar condition as an indication that only quantizing the nuclear variables while keeping electronic states classical is only a good approximation when the problem involves ``flat" diabatic potential matrices uncoupled from the nuclear dofs.  With this approximation, $\bm{\Delta M}_{ee}(t) =0$ and following similar reasoning as in NHWD(E) and HWD(A), the electronic prefactor cancels the Hessian from the SPA. The equations of motion for the nuclear monodromy matrices are, 
\begin{align}
    \dot{\bm{\bar{M}}}_{NN} & = \mathbb{A}_{NN}\cdot\bm{\bar{M}}_{NN} + \mathbb{B}_{NN}\cdot\bm{\Delta M}_{NN}, \\ 
    \bm{\Delta}\dot{\bm{M}}_{NN} & = \mathbb{A}_{NN}\cdot\bm{\Delta M}_{NN} + \frac{1}{4}\mathbb{B}_{NN}\cdot\bm{\bar{M}}_{NN},
\end{align}
with,
\begin{align}
    \mathbb{A}_{NN} &= \left(\begin{array}{cc}
    \mathbb{0} & \bm{m}^{-1} \\
    -\frac{1}{2}\mathrm{h}_{+}^{\prime\prime}\left(\bm{\bar{z}},\bm{X}^{\pm}\right)  &  \mathbb{0} \\ 
    \end{array}\right), \\
    \mathbb{B}_{NN} &= \left(\begin{array}{cc}
    \mathbb{0} & \mathbb{0} \\
    -\mathrm{h}_{-}^{\prime\prime}\left(\bm{\bar{z}},\bm{X}^{\pm}\right)  &  \mathbb{0} \\ 
    \end{array}\right),
\end{align}
initial conditions, $\bm{\bar{M}}_{NN}(0) = \mathbb{1}$, $\bm{\Delta M}_{NN}(0) = 0$. The monodromy matrices satisfy symplecticity conditions similar to Eqs.~\eqref{eq:sym_mb_dm1} and~\eqref{eq:sym_mb_dm2}, with $\bm{J}$ replaced by $\bm{J}_N$.
The NHWD(N) approximation to a correlation function becomes, 
\begin{align}
    C_{AB}^{\text{NHWD(N)}}(t) & = \frac{1}{\left(2\pi\hbar\right)^{2D + F}} \int d\bm{Z}_{0}^{\pm} \int d\bm{\bar{z}}_{0} \, \Tilde{{\rho}}_{A_N}^{*}(\bm{Z}_{0}^{\pm})\left[\hat{\rho}_{A_e}\right]_W\left(\bm{\bar{z}}_{0}\right) \notag \\
    & \times   \Tilde{B_N}(\bm{Z}_{t}^{\pm}) \left[\hat{B}_e\right]_W\left(\bm{\bar{z}}_{t}\right) \tilde{\mathcal{C}}_{t}^{\text{NHWD(N)}}\left(\bm{Z}_{0}^{\pm},\bm{\bar{z}}_{0}\right) \notag \\  
& \times e^{i\tilde{S}_t^{\text{NHWD(N)}}\left(\bm{Z}_{0}^{\pm},\bm{\bar{z}}_{0}\right)/\hbar} .
\end{align}
Again we have assumed that the operator $\hat{B}\equiv \hat{B}_e \otimes \hat{B}_N$ factorizes into nuclear and electronic parts and similarly for $\hat{\rho}_{A}$. The nuclear prefactor is, 
\begin{align}  
  & \tilde{\mathcal{C}}_{t}^{NHWD(N)}\left(\bm{Z}_{0}^{\pm},\bm{\bar{z}}_{0}\right) \notag \\
  & = \text{det}\left(2\bm{\Gamma}_N\right)^{-1/2}\text{det}\left[\bm{\Gamma}_N\cdot\bm{\bar{M}}_{NN} - \bm{J}_N\cdot\bm{\bar{M}}_{NN}\cdot\bm{J}_N\cdot\bm{\Gamma}_N  \right. \notag  \\
    & \left. + 2i\bm{J}_N\cdot\bm{\Delta M}_{NN}   + 2i\bm{\Gamma}_N\cdot \bm{\Delta M}_{NN}\cdot\bm{J}_N\cdot\bm{\Gamma}_N  \right]^{1/2},
\end{align}
and the action is,
\begin{align}
   &  \tilde{S}_t^{NHWD(N)}(\bm{Z}^{\pm}_0,\bm{\bar{z}}_0) \notag \\
    &  \equiv -\bm{\bar{P}}^{T}_t\cdot\bm{\Delta X}_t + \bm{\bar{P}}^{T}_0\cdot\bm{\Delta X}_0 + \tilde{S}_t(\bm{Z}^{+}_0,\bm{\bar{z}}_0) - \tilde{S}_t(\bm{Z}^{-}_0,\bm{\bar{z}}_0) \notag \\ 
    & = -\bm{\bar{P}}^{T}_t\cdot\bm{\Delta X}_t + \bm{\bar{P}}^{T}_0\cdot\bm{\Delta X}_0 \notag \\
     & + \int_0^{t} d\tau \, \left[ \bm{\bar{P}}_{\tau}\cdot\bm{m}^{-1}\cdot\bm{\Delta P}_{\tau} - \left\{\mathrm{h}_{el}(\bm{\bar{z}}_{\tau},\bm{X}^{+}_{\tau}) - \mathrm{h}_{el}(\bm{\bar{z}}_{\tau},\bm{X}^{-}_{\tau}) \right\} \right]. \label{eq:act_HWDn}
\end{align}
NHWD(N) will primarily be used when nuclear quantum effects are necessary to include. However, the accuracy of this method is untested as of now, and the fact that electronic states are being treated classical may render this method inaccurate.

\section{Deriving NHWD(V)} \label{app:nhwdv}

To derive NHWD(V), we start from the full DHK-Wigner correlation function, Eq.~\eqref{eq:cab_dhk_mmst} and perform the integrals over the nuclear bath $\bm{\Xi}^T_{b} = \left(\bm{Z}_b,\bm{Z}^{\prime}_b,\bm{\Delta Z}_{0,b}\right)^T$ by SPA while keeping the integrals over the vibronic system   $\bm{\xi}_{s} = \left(\bm{\mathsf{Z}}_{s},\bm{\mathsf{Z}}^{\prime}_{s},\bm{\Delta \mathsf{Z}}_{0,s}\right)^T$, exact. The integrals over all the mean variables $\bm{\bar{\mathsf{Z}}}_0$ are performed exactly, and the phase in the SPA is the same as in Eq.~\eqref{eq:phi}. Again, as can be expected by now, the the stationary phase conditions emerge as,
\begin{align}
    \bm{Z}_b &= \bm{\bar{Z}}_{0,b}, & \bm{Z}^{\prime}_b &= \bm{\bar{Z}}_{t,b},  & \text{and} &  & \bm{\Delta Z}_{t,b} &= 0. \label{eq:HWD_v_sp_conditions} 
\end{align}
As in the other cases, these are not sufficient to satisfy all three SP equations. We must make a further approximation that the system-bath and bath-system blocks of the monodromy matrices vanish to satisfy all three SP equations. Following similar analysis as in the other cases, this can be achieved if the following blocks of matrices $\mathbb{A}$ and $\mathbb{B}$ vanish:
\allowdisplaybreaks
\begin{align}
    \mathbb{A}_{sb} & = \left(\begin{array}{cc}
    \frac{\partial\dot{\bm{\bar{x}}}}{\partial \bm{\bar{X}}_b} &  \mathbb{0} \\
    \mathbb{0} & \mathbb{0} \\
    - \frac{\partial\dot{\bm{\bar{p}}}}{\partial \bm{\bar{X}}_b} &  \mathbb{0} \\
    - \frac{\partial^2 \mathrm{h}_+}{\partial \bm{\bar{X}}_b\partial\bm{\bar{X}}_s}  &  \mathbb{0}
    \end{array}\right), \notag \\  
    \mathbb{A}_{bs} & = \left(\begin{array}{cccc}
\mathbb{0} &  \mathbb{0} &  \mathbb{0} &  \mathbb{0}  \\
- \frac{\partial\dot{\bm{\bar{p}}}}{\partial \bm{\bar{X}}_b}^T & - \frac{\partial^2 \mathrm{h}_+}{\partial \bm{\bar{X}}_s\partial\bm{\bar{X}}_b} & - \frac{\partial\dot{\bm{\bar{x}}}}{\partial \bm{\bar{X}}_b}^T & \mathbb{0}  \\
    \end{array}\right), \notag \\  
    \mathbb{B}_{sb} & = 4\left(\begin{array}{cc}
    \frac{\partial\dot{\bm{\bar{x}}}}{\partial \bm{\Delta {X}}_b} &  \mathbb{0} \\
    \mathbb{0} & \mathbb{0} \\
    - \frac{\partial\dot{\bm{\bar{p}}}}{\partial \bm{\Delta {X}}_b} &  \mathbb{0} \\
    - \frac{\partial^2 \mathrm{h}_-}{\partial \bm{\bar{X}}_b\partial\bm{\bar{X}}_s}  &  \mathbb{0}
    \end{array}\right), \notag \\
    \mathbb{B}_{bs} & = \left(\begin{array}{cccc}
      \mathbb{0} &  \mathbb{0} & \mathbb{0} &  \mathbb{0} \\
    \frac{\partial\bm{\Delta}\dot{\bm{{p}}}}{\partial \bm{\bar{X}}_b}^T & - \frac{\partial^2 \mathrm{h}_-}{\partial \bm{\bar{X}}_s\partial\bm{\bar{X}}_b} & -\frac{\partial\bm{\Delta}\dot{\bm{{x}}}}{\partial \bm{\bar{X}}_b}^T & \mathbb{0} \\
    \end{array}\right), \label{eq:ab_vib}
\end{align}
where it is understood that the Hessians of  $\mathrm{h}_{\pm}$  are evaluated with the SP condition imposed. Setting these to zero can be interpreted as the forces on the vibronic system being independent of the nuclear bath. In the case where the nuclear modes in the vibronic system are uncoupled from those in the bath, the relevant parts of the Hessians of  $\mathrm{h}_{\pm}$ are zero, but this is generally not the case. 

With these assumptions, all three stationary phase equations are satisfied, and the resulting trajectories conserve the Hamiltonian, Eq.~\eqref{eq:nhwdv_ham} and follow the equations of motion, Eqs.~\eqref{eq:nhwdv_eom1}-\eqref{eq:nhwdv_eom2}. The vibronic system has forward and backward trajectories that feel the mean trajectory of the bath. The assumption of the vanishing off-diagonal monodromy matrix blocks separates the prefactor into vibronic system and nuclear bath prefactors. The components of electronic Hamiltonian take the form, 
\begin{align}
    U(\bm{X}) & \equiv U_b(\bm{X}_b) + U_{sb}(\bm{X}) + U_s(\bm{X}_s) \notag \\ 
    \bm{V}(\bm{X}) & \equiv  \bm{V}_s(\bm{X}_s) +  \bm{V}_{sb}(\bm{X}) +  \bm{V}_b(\bm{X}_b) \label{eq:pot-split}
\end{align}
where the  former are state-independent potential terms for the bath, system-bath coupling and the system respectively, whereas the later are the system part, the system-bath coupling and bath part of the state-dependent diabatic potential energy matrix. For such a Hamiltonian, one of the assumptions required to make the off-diagonal blocks of the Monodromy matrix vanish, amounts to setting $U_{sb}(\bm{X}) = \bm{V}_{sb}(\bm{X}) = 0 $ in the monodromy matrix propagation. This is equivalent to setting $\frac{\partial^2 \mathrm{h}_{\pm}}{\partial \bm{\bar{X}}_s\partial\bm{\bar{X}}_b} = 0$ in Eq.~\eqref{eq:ab_vib}. For many standard models mentioned in the main text, $\bm{V}_b^{\prime\prime}(\bm{X}_b) = 0$,  and this approximation also implies $\frac{\partial^2 \mathrm{h}_{-}}{\partial \bm{\bar{X}}_b^2} = 0$, in turn implying $\bm{\Delta M}_{bb}(t)=0$. With this approximation, the bath prefactor cancels the Hessian due to the SPA approximation, following the arguments from the AHWD case in paper I.\cite{Malpathak2024b} We thus only need to propagate the system monodromy matrix, which follows the equations of motion, 
\begin{align}
    \dot{\bm{M}}^{\pm}_{ss}= \bm{J}_s\cdot\frac{\partial^2 \bar{H}(\bm{\mathsf{Z}}_s^{\pm},\bm{\bar{Z}}_b)}{\partial \bm{\mathsf{Z}}^{\pm^2}_s}\cdot\bm{M}^{\pm}_{ss}, \label{eq:nhwdv_mono_eom}
\end{align}
with $\bm{M}_{ss}^{\pm}(0)=\mathbb{1}$, and follows the symplecticity condition
\begin{align}                   
\bm{M}_{ss}^{{\pm}^T}\cdot\bm{J}_{s}\cdot\bm{M}^{\pm}_{ss}=\bm{J}_{s}. \label{eq:nhwdv_symp}
\end{align}
Putting all the pieces together the expression for the NHWD(V) correlation function is,
\begin{align}
    C_{AB}^{\text{NHWD(V)}}(t) & = \frac{1}{\left(2\pi\hbar\right)^{2N_s + N_b}} \int d\bm{\mathsf{Z}}_{0,s}^{\pm} \int d\bm{\bar{Z}}_{0,b} \, \Tilde{{\rho}}_{A_s}^{*}(\bm{\mathsf{Z}}_{0,s}^{\pm}) \Tilde{B_s}(\bm{\mathsf{Z}}_{t,s}^{\pm}) \notag \\
    & \times \left[\hat{\rho}_{A_b}\right]_W\left(\bm{\bar{Z}}_{0,b}\right)   \left[\hat{B}_b\right]_W\left(\bm{\bar{Z}}_{t,b}\right) \notag \\  & \times \tilde{\mathcal{C}}_{t}^{\text{NHWD(V)}}(\bm{\mathsf{Z}}_{0,s}^{\pm},\bm{\bar{Z}}_{0,b})e^{i\tilde{S}_t^{\text{NHWD(V)}}(\bm{\mathsf{Z}}_{0,s}^{\pm},\bm{\bar{Z}}_{0,b})/\hbar}.
\end{align}
The NHWD(V) prefactor only requires the system-system block of the monodromy matrix, and is defined as, 
\begin{align}
    & \tilde{\mathcal{C}}_{t}^{\text{NHWD(V)}}(\bm{\mathsf{Z}}_{0,s}^{\pm},\bm{\bar{Z}}_{0,b}) \notag \\
    & =  \text{det}\left(2\bm{\Gamma}_s\right)^{-1/2}  \text{det}\left[\frac{1}{2}\left(\bm{\Gamma}_s+i\bm{J}_s\right)\cdot\bm{M}_{ss}^+\cdot\left(\mathbb{1}_s+i\bm{J}_s\cdot\bm{\Gamma}_s\right) \right. \notag \\
    & \left. + \frac{1}{2}\left(\bm{\Gamma}_s-i\bm{J}_s\right)\cdot\bm{M}_{ss}^-\cdot\left(\mathbb{1}_s-i\bm{J}_s\cdot\bm{\Gamma}_s\right)\right]^{1/2}. \label{eq:nhwdv_pref}
\end{align}
The NHWD(V) action is, 
\begin{align}
   & \tilde{S}_t^{NHWD(V)}(\bm{\mathsf{Z}}^{\pm}_{0,s},\bm{\bar{Z}}_{0,b}) \notag \\ 
   &  \equiv -\bm{\bar{\mathsf{P}}}^{T}_{t,s}\cdot\bm{\Delta \mathsf{X}}_{t,s} + \bm{\bar{\mathsf{P}}}^{T}_{0,s}\cdot\bm{\Delta \mathsf{X}}_{0,s}  + \tilde{S}_t(\bm{\mathsf{Z}}^{+}_{0,s},\bm{\bar{Z}}_{0,b}) - \tilde{S}_t(\bm{\mathsf{Z}}^{-}_{0,s},\bm{\bar{Z}}_{0,b}) \notag \\
   & = -\bm{\bar{\mathsf{P}}}^{T}_{t,s}\cdot\bm{\Delta \mathsf{X}}_{t,s} + \bm{\bar{\mathsf{P}}}^{T}_{0,s}\cdot\bm{\Delta \mathsf{X}}_{0,s} + \int_0^{t} d\tau \, \left[ \bm{\bar{P}}_{\tau,s}\cdot\bm{m}^{-1}_s\cdot\bm{\Delta P}_{\tau,s} \right. \notag \\
   & + \left. \dot{\bm{x}}^{+}_{\tau}\bm{p}^{+}_{\tau} -  \dot{\bm{x}}^{-}_{\tau}\bm{p}^{-}_{\tau} - \left\{\mathrm{h}_{el}(\bm{\mathsf{Z}}^{+}_{\tau,s},\bm{\bar{X}}_{\tau,b}) - \mathrm{h}_{el}(\bm{\mathsf{Z}}^{-}_{\tau,s},\bm{\bar{X}}_{\tau,b})  \right\}\right]. \label{eq:act_HWDv}
\end{align}
Details of the correlation function, trajectory structure and the prefactor are provided in the main text.

\begin{figure}
    \centering
    \includegraphics[width=0.45\textwidth]{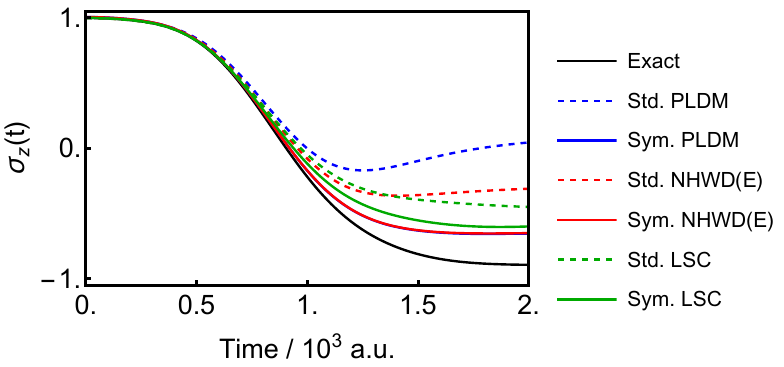}
    \caption{ Expectation values as a function of time are plotted for the Pauli spin matrix $\bm{\sigma}_z$ for low energy scattering model. Calculations performed using LSC (green), NHWD(E) (red), and PLDM (blue) are compared against exact results. Solid and dashed lines denote results calculated using the symmetrized and standard MMST Hamiltonian respectively.}
    \label{fig:std_vs_sym}
\end{figure}

\section{Standard vs. Symmetrized MMST Hamiltonian} \label{app:comp_hams}
In this appendix we present calculations of $\langle\bm{\sigma}_z\rangle$ for the low energy case of scattering model 2 predicted by various methods, when the symmetrized versus the conventional MMST mapping Hamiltonian is used. As seen in Fig.~\ref{fig:std_vs_sym} and has been reported earlier,\cite{Kelly2012} the symmetrized version yields more accurate results for all methods studied here. More interestingly, NHWD(E) and PLDM results are identical with the symmetrized Hamiltonian, but that is not the case with the conventional Hamiltonian. This is also observed for the high-energy case and scattering model 1 (not shown). This observation seems to suggest that in general NHWD(E) is a distinct method when compared to PLDM, as also evidenced by other differences mentioned in the main text. However, when a symmetrized Hamiltonian is used, there might be connections between the two. These will be investigated in future studies. 

\bibliographystyle{apsrev4-2}
\bibliography{bibfile}

\end{document}